\documentclass[11pt]{article}
\pdfoutput=1 

\usepackage{jcappub} 

\usepackage[T1]{fontenc} 
\usepackage{braket}
\usepackage{xcolor}

\begin{document}

\title{Bound States of Pseudo-Dirac Dark Matter}

\author[a]{Arindam Bhattacharya,}
\author[a,b]{Tracy R. Slatyer}
\affiliation[a]{Center for Theoretical Physics\\Massachusetts Institute of Technology, Cambridge, MA 02139, USA}
\affiliation[b]{School of Natural Sciences\\Institute for Advanced Study, Princeton, NJ 08540, USA}

\emailAdd{arindamb@mit.edu}
\emailAdd{tslatyer@mit.edu}

\abstract{We study the bound-state spectrum in a simple model of pseudo-Dirac dark matter, and examine how the rate of bound-state formation through radiative capture compares to Sommerfeld-enhanced annihilation. We use this model as an example to delineate the new features induced by the presence of a mass splitting between the dark matter and a nearly-degenerate partner, compared to the case where only a single dark-matter-like state is present. We provide a simple analytic prescription for estimating the spectrum of bound states in systems containing a mass splitting, which in turn allows characterization of the resonances due to near-zero-energy bound states, and validate this estimate both for pseudo-Dirac dark matter and for the more complex case of wino dark matter. We demonstrate that for pseudo-Dirac dark matter the capture rate into deeply bound states is, to a good approximation, simply related to the Sommerfeld enhancement factor.}

\preprint{MIT-CTP/5091}

\maketitle

\section{Introduction}
\label{sec:intro}

In recent years there has been considerable interest in the possibility that dark matter (DM) could form bound states, which are ubiquitous in the Standard Model of particle physics. Such bound states could be a consequence of weakly-coupled interactions between the DM and a light mediator (e.g. \cite{Pospelov:2008jd,Kaplan:2009de,Wise:2014jva,Gresham:2017zqi}), or alternatively a strongly-interacting dark sector (e.g. \cite{Frigerio:2012uc,Detmold:2014qqa,Detmold:2014kba,Francis:2018xjd}). For sufficiently heavy dark matter, even interactions through the electroweak gauge bosons are sufficient to support bound states (e.g. \cite{Asadi:2016ybp, Mitridate:2017izz}).

The presence of bound states could lead to novel signatures across a wide range of observational probes, including colliders \cite{Drees:1993uw,Martin:2008sv,Shepherd:2009sa,Kats:2009bv,Kats:2012ym,An:2015pva,Tsai:2015ugz,Elor:2018xku,Krovi:2018fdr} and direct-detection experiments (e.g. \cite{Kaplan:2009de,Laha:2013gva}). In indirect detection, formation of unstable bound states constitutes an additional annihilation channel for the DM, which in some circumstances can dominate over direct annihilation \cite{an2016effects,an2017strong,cirelli2017dark}. With sufficiently good experimental energy resolution, the decays of bound states could be distinguished from direct annihilation; the soft particles radiated in transitions into the bound state, and between bound states in the spectrum, could also lead to observable signals \cite{MarchRussell:2008tu}. Formation of bound states could modify the cosmological history of DM (e.g. \cite{vonHarling:2014kha,Wise:2014jva}), and if the bound states are stable, their presence could also have astrophysical effects in the late universe (e.g. \cite{Fan:2013yva,Gresham:2018anj}). Finally, the presence of bound states is connected to the existence of DM self-interaction, which could have striking effects on the distribution of dark matter at Galactic scales (see \cite{Buckley:2017ijx} for a recent review).

One simple illustrative model for self-interacting DM is where the DM is a member of a multiplet charged under some dark gauge group, with a small breaking of the gauge symmetry conferring a small mass on the corresponding gauge boson. The breaking of the symmetry relating the different components of the multiplet will also generically lead to a mass splitting between those components, with the lightest playing the role of the DM. This scenario is realized in the case of wino or higgsino DM \cite{jungman1996supersymmetric}, and extensions to the case of higher representations of the electroweak gauge group \cite{Cirelli:2005uq} or simple dark sectors (e.g. \cite{ArkaniHamed:2008qn, Cheung:2009qd, Cirelli:2010nh}) have been studied in the literature.

DM annihilation and self-interaction in such models has been studied previously, taking into account the presence of the mass splittings as well as the long-range interaction \cite{Hisano:2003ec,hisano2005nonperturbative,slatyer2010sommerfeld, schutz2015self, Zhang:2016dck, Vogelsberger:2018bok}. In this work we will often refer to the Sommerfeld enhancement, by which we simply mean the enhancement of short-range annihilation processes due to the long-range potential from vector exchange. However, our current analytic understanding of the bound states in such scenarios is largely limited to the case without mass splittings \cite{Petraki:2015hla, Asadi:2016ybp, Petraki:2016cnz, Mitridate:2017izz, Harz:2018csl}; where mass splittings are present, previous studies have relied on numerical work.

In this work, we seek to address this gap. We consider a simple low-energy scenario containing two nearly-degenerate Majorana fermions interacting through vector exchange. The lighter fermion is the DM, the heavier can be thought of as an excited state of the DM. We calculate numerically how the mass splitting between the states alters the bound state spectrum and capture rate relative to the case with only a single state, and develop a simple analytic understanding of the main effects. If dark matter is a Majorana fermion, the interaction with the gauge bosons must be off-diagonal in nature as the DM cannot carry a conserved charge \cite{slatyer2010sommerfeld,schutz2015self}; this setup is naturally realized where the DM is a Dirac fermion charged under the dark gauge symmetry at high energies, and separates into two nearly-degenerate Majorana fermions at low energies due to the symmetry breaking (see e.g. \cite{Finkbeiner:2010sm} for a detailed discussion). Previous studies of this simple model \cite{slatyer2010sommerfeld,schutz2015self} have considered $s$-wave annihilation and scattering; we go beyond by considering bound state formation, and including higher partial waves as well.

We demonstrate that we can analytically estimate the shift to the bound state energies in the presence of a mass splitting, and identify regions of parameter space where mass-splitting-induced changes to the capture cross section follow characteristic patterns. We show that the changes to the capture cross section are dominated by the behavior of the initial-state wavefunction, and the resulting cross section is simply related to the Sommerfeld-enhanced direct annihilation rate (for the same partial wave in the initial state) up to a phase-space factor, to a good approximation. We apply our understanding from this simple model to the case of wino dark matter, and demonstrate that our analytic approximations to the bound state energies in the latter case compare well to previous numerical results \cite{Asadi:2016ybp}. 

In Section \ref{sec:toy}, we describe the pseudo-Dirac dark matter model,  lay out the relevant non-relativistic Hamiltonian for the problem, and discuss its general properties. In Section \ref{sec:bound}, we discuss the general structure of the bound state spectrum, describe our method for numerically obtaining the bound-state energies, provide an analytic estimate for the shift in bound state energies as a function of the mass splitting, and compare analytic and numerical results for the binding energies and the effect on the position of resonances in the scattering rate. In Section \ref{sec:wino} we apply our analytical insights from this toy model to the case of the wino, as a test case for bound states of electroweakly interacting DM. In Section \ref{sec:capture} we numerically compute the capture rate in this model, and then characterize and discuss the new features relative to the case with no mass splitting, in particular relating the capture rate to the Sommerfeld enhancement. We summarize our conclusions in Section~\ref{sec:conclusion}.

\section{Pseudo-Dirac dark matter: general considerations}\label{sec:toy}
\subsection{The model}
We consider a pseudo-Dirac fermion, charged under a $U(1)$ gauge group in the dark sector, which acquires a Majorana mass term at low energies due to the breaking of the $U(1)$ symmetry by an Abelian dark Higgs. Consequently, the Dirac fermion splits into two non-degenerate Majorana fermion mass eigenstates. We will be solely interested in the low-energy, long-range behavior of the system in the  non-relativistic limit, encoded in the potential; consequently, there is considerable freedom in the details of the dark Higgs sector. For example, if the dark Higgs has (dark $U(1)$) charge 1, a small Majorana mass term can be induced by a dimension-5 operator \cite{Finkbeiner:2010sm}; if the dark Higgs instead has charge 2, a Yukawa-type coupling yields a Majorana mass when the Higgs acquires a VEV, as in \cite{Elor:2018xku} (in this case the small splitting between eigenstates may be due to a small Dirac mass, as the Majorana mass is not suppressed by a high scale). We label the lower and upper mass eigenstates as $\chi$ and $\chi^*$ respectively, and hereby denote the mass eigenvalues as $(m_{\chi}, m_\chi + 2 \delta)$. 

More explicitly, at high energies, writing the Dirac fermion as $\Psi$ and its Dirac mass as $m_D$, and the gauge boson of the $U(1)$ symmetry as $\phi$, the Lagrangian can be written as:
\begin{equation} \mathcal{L} = i \bar{\Psi} \gamma^\mu (\partial_\mu - i g_D \phi_\mu) \Psi - m_D \bar{\Psi} \Psi + \mathcal{L}_\text{Higgs} + \mathcal{L}_\text{gauge-kin},\end{equation}
where $g_D$ is the dark-sector coupling, and we have omitted the details of the model-dependent Higgs sector. The gauge kinetic term is by default just $\mathcal{L}_\text{gauge-kin} = -\frac{1}{4} F^D_{\mu \nu} F^{D\mu \nu}$, where $F^D_{\mu \nu}$ is the dark field strength associated with the $\phi^\mu$ field, but it could also include e.g. kinetic mixing with the Standard Model gauge bosons. Writing $\Psi$ as a Weyl fermion pair $(\zeta,\eta^\dagger)$ (see Ref.~\cite{Dreiner:2008tw} for an extended discussion), turning on a Higgs-sector-induced Majorana mass $m_M$ for $\Psi$ yields a mass matrix of the form \cite{Finkbeiner:2010sm,Elor:2018xku}:
\begin{equation} \frac{1}{2} \begin{pmatrix} \zeta & \eta \end{pmatrix} \begin{pmatrix} m_M & m_D \\ m_D & m_M\end{pmatrix} \begin{pmatrix} \zeta \\ \eta \end{pmatrix} + h.c.\end{equation}
This leads to mass eigenstates that are a 45$^\circ$ rotation of $\zeta$, $\eta$, i.e. $\chi^* = (\eta + \zeta)/\sqrt{2}$, with mass $m_M + m_D$, and $\chi = i (\eta - \zeta)/\sqrt{2}$, with mass $|m_M - m_D|$. This rotation converts the gauge boson-fermion interaction $g_D \bar{\Psi} \gamma^\mu \phi_\mu \Psi$ into the form \cite{Elor:2018xku}:
\begin{equation} \mathcal{L}_\text{fermion-gauge} = - i g_D \phi_\mu \left((\chi^*)^\dagger\bar{\sigma}^\mu \chi - \chi^\dagger \bar{\sigma}^\mu \chi^*\right).\end{equation}

We observe that the interaction between the mass eigenstates via $\phi_\mu$ is off-diagonal in nature, i.e. it couples the $\chi$ and $\chi^*$ fields. Since Majorana fermions cannot carry conserved charge, this off-diagonal interaction is generic for any system where two Majorana fermions interact through vector exchange. 

This off-diagonal interaction structure gives rise to two distinct sectors of two-body states comprised of $\chi$ and $\chi^*$. $\ket{\chi\chi^*}$ states are maintained (converted into the identical state $\ket{\chi^*\chi}$) under the exchange of a vector boson. In contrast, such a vector exchange transforms $\ket{\chi\chi}$ states into $\ket{\chi^*\chi^*}$; thus the potential mixes the non-degenerate $\ket{\chi\chi}$ and $\ket{\chi^*\chi^*}$ states. The $\ket{\chi\chi^*}$ states experience a simple attractive Yukawa potential in the non-relativistic limit, whereas the admixed $\alpha \ket{\chi\chi} + \beta \ket{\chi^*\chi^*}$ states evolve under a more complex potential.

For the purposes of our study, we will be primarily interested in the dynamics of the two-state system spanned by $\ket{\chi\chi}$ and $\ket{\chi^{*}\chi^{*}}$, and the effect of the mass splitting between these states. In the non-relativistic limit, the Schr\"{o}dinger equation for this system contains a matrix potential of the following form \cite{arkani2009theory}:
\begin{equation}\label{potential}
    V(r)=\begin{bmatrix}
    0 & -\hbar c\alpha\frac{e^{-m_{\phi}cr/\hbar}}{r} \\
    -\hbar c\alpha\frac{e^{-m_{\phi}cr/\hbar}}{r} & 2\delta c^2
    \end{bmatrix}
\end{equation}where $\alpha$ denotes the dark coupling constant $g_D^2/4\pi$, $m_\phi$ is the mass of the dark gauge boson, and $r$ is the inter-particle separation. The first row (and column) corresponds to the two-body $\ket{\chi \chi}$ state and the second row (and column) to the two-body $\ket{\chi^*\chi^*}$ state. The off-diagonal terms represent the conversion of $\ket{\chi \chi}$ into $\ket{\chi^*\chi^*}$, and vice versa, via the vector exchange, while the $2\delta$ term describes the increased mass of the $\ket{\chi^*\chi^*}$ state.

As in Ref.~\cite{slatyer2010sommerfeld}, we scale the coordinate $r$ by $\frac{\hbar}{m_{\chi}\alpha c}$, thereby obtaining the following radial equation for the reduced wavefunction $\psi(r)$ in the centre-of-mass frame, 
\begin{equation}\label{matrix}
    \psi''(r)=\begin{bmatrix}
    \frac{l(l+1)}{r^2}-\epsilon_{v}^2 & -\frac{e^{-\epsilon_{\phi}r}}{r}\\
    -\frac{e^{-\epsilon_{\phi}r}}{r} & \epsilon_{\delta}^2+\frac{l(l+1)}{r^2}-\epsilon_{v}^2
    \end{bmatrix}\psi(r)
\end{equation}with the dimensionless parameters defined as 
\begin{align}\label{eq:dimensionlessparams}
    \epsilon_{v}&=\frac{v}{c\alpha}& 
    \epsilon_{\phi}&=\frac{m_{\phi}}{m_{\chi}\alpha}&
    \epsilon_{\delta}=\sqrt{\frac{2\delta}{m_{\chi}\alpha^2}}
\end{align}

\subsection{Eigenvectors and eigenvalues of the potential}\label{sec:general considerations}

Eq.~\ref{matrix} cannot be solved analytically in general, since the diagonalizing matrix is itself position-dependent. However, it is still helpful to examine the eigenvalues and eigenvectors of the potential, which are respectively given by \cite{slatyer2010sommerfeld,schutz2015self}:
\begin{equation}\label{eigenvectors}
    \lambda_{\pm}=-\epsilon_{v}^2+\frac{l(l+1)}{r^2}+\frac{\epsilon_{\delta}^2}{2}\pm\sqrt{\bigg(\frac{\epsilon_{\delta}^2}{2}\bigg)^2+\bigg(\frac{e^{-\epsilon_{\phi}r}}{r}\bigg)^2}\text{,}~\eta_{\pm}(r)=\frac{1}{\sqrt{2}}\begin{bmatrix}
    \mp\sqrt{1\mp\frac{1}{\sqrt{1+\big(\frac{2e^{-\epsilon_{\phi}r}}{r\epsilon_{\delta}^2}\big)^2}}}\\
    \sqrt{1\pm\frac{1}{\sqrt{1+\big(\frac{2e^{-\epsilon_{\phi}r}}{r\epsilon_{\delta}^2}\big)^2}}}
    \end{bmatrix}
\end{equation}

The expressions in Eq.~\ref{eigenvectors} allow the identification of two interesting regimes, where the eigenvectors are nearly $r$-independent and the problem is approximately diagonalizable:\begin{itemize}
    \item \textbf{Small $r$ regime}: For sufficiently small $r$, the Yukawa potential dominates the mass-splitting term, yielding: \begin{equation}\label{smallr}
        \lambda_{\pm}\approx-\epsilon_{v}^2+\frac{l(l+1)}{r^2}+\frac{\epsilon_{\delta}^2}{2}\pm\frac{e^{-\epsilon_{\phi}r}}{r}+\mathcal{O}(r),~\eta_{\pm}(r)\approx\frac{1}{\sqrt{2}}\begin{bmatrix}
        \mp1\\1\end{bmatrix}+\mathcal{O}(r)
    \end{equation}The eigenvectors of the potential are those of $V(r)=-\frac{1}{r}\sigma_{x}$, and the eigenvalues physically correspond to the repulsive and attractive potential appropriate to same-sign or opposite-sign scattering respectively. This regime corresponds to the restoration of the $U(1)$ symmetry in the ultraviolet.
    \item\textbf{Large $r$ limit}: Far away from the origin, one observes that the mass splitting term dominates the now weak Yukawa potential, leading to \begin{align}\label{adia}
        \lambda_{+}&\approx-\epsilon_{v}^2+\frac{l(l+1)}{r^2}+\epsilon_{\delta}^2+\frac{e^{-2\epsilon_{\phi}r}}{\epsilon_{\delta}^2r^2}+\mathcal{O}\bigg(\frac{1}{r^4}\bigg),\quad\quad \eta_{+}(r)\approx\begin{bmatrix}0\\1\end{bmatrix}+\mathcal{O}\bigg(\frac{1}{r}\bigg)\\
        \lambda_{-}&\approx-\epsilon_{v}^2+\frac{l(l+1)}{r^2}-\frac{e^{-2\epsilon_{\phi}r}}{\epsilon_{\delta}^2r^2}+\mathcal{O}\bigg(\frac{1}{r^4}\bigg),\quad\quad \eta_{-}(r)\approx\begin{bmatrix}1\\0\end{bmatrix}+\mathcal{O}\bigg(\frac{1}{r}\bigg)
        \end{align}
\end{itemize}
Thus, we find that the eigenvectors undergo a rotation at a radius where the Yukawa potential and the mass splitting are comparable in size, as also noted in Ref.~\cite{slatyer2010sommerfeld}. 

For bound states where the support of the wavefunction lies primarily within this radius, we can guess that the potential in the small-$r$ regime will be a reasonable approximation when computing the bound-state spectrum, allowing us to ignore the radial variation of the eigenstates/eigenvalues.

\section{The bound-state spectrum}\label{sec:bound}

As discussed previously, there are two types of bound states supported by the dark vector exchange, $\ket{\chi\chi^*}$ states which are supported by a simple Yukawa potential, and states consisting of an admixture of $\ket{\chi\chi}$, $\ket{\chi^*\chi^*}$, which evolve under the potential given in Eq.~\ref{potential}. The admixed states consist of pairs of identical fermions, and thus must be in an antisymmetric configuration; this corresponds to requiring the sum of their orbital and spin angular momentum quantum numbers $L+S$ to be even (see Ref.~\cite{Beneke:2014gja} for a general discussion of the different potentials experienced by even- and odd-$L+S$ states).

These bound states can be produced by radiative capture from scattering states. The dominant contribution to such processes arise from  electric-dipole-like transitions with emission of a single particle. These transitions change the angular momentum of the incoming state by $\pm1$, if a vector particle is emitted (such as a photon or dark photon). This restricts the possible types of transitions. For our purposes, we shall assume the scattering state to be $\chi\chi$ at large enough interparticle separation (since only $\chi$ particles are present in the DM halo, unless the mass splitting is small enough to be comparable to the typical DM kinetic energy). The capture is then into bound states in the $\chi\chi^*$ sector, via the emission of a single dark photon. Since the initial wavefunction must have even $L+S$, the bound state in this case has odd $L+S$; for example, the $s$-wave ($L=0$) contribution to the scattering state only receives contributions from spin-singlet states ($S=0$), and capture could occur into an $np$ spin-singlet bound state ($L=1$,$S=0$) where $n\geq 2$. Similarly, to form a spin-triplet bound state by direct radiative capture, the dominant process is capture into $ns$ states with $n \geq 1$, from the $p$-wave or $d$-wave components of the initial state.

The presence of bound states in the $\ket{\chi\chi}+\ket{\chi^*\chi^*}$ sector, with energies close to zero, can lead to resonant enhancement to the radiative capture cross section at low velocities. When there is a bound state with near-zero energy present in the spectrum of $L+S$-even states, the scattering wavefunction for $L+S$-even states is enhanced at short distances, leading to the resonance peaks in the Sommerfeld enhancement \cite{Hisano:2003ec} and also enhancing the radiative capture rate into the $L+S$-odd bound states. The energies of these bound states, and thus the conditions under which resonances occur, depend both on the force carrier mass $m_\phi$, and on the mass splitting between states $\delta$. 

While analytic solutions do not exist for bound states in an attractive Yukawa potential, they closely resemble bound states in the Hulth\'{e}n potential, which is given by\begin{equation}
    V_\text{H}(r)=-\frac{\alpha_{H} m_{\text{H}}}{e^{m_{\text{H}}r}-1}
\end{equation}where $\alpha_{H}$ is the relevant coupling and $m_{H}$ characterizes the range of the potential. The Hulth\'{e}n potential has the desired asymptotic behaviour, resembling a Coulomb potential in the low-$r$ limit and decreasing exponentially in the large-$r$ limit, similar to the Yukawa potential. Exact solutions for the $s$-wave states and approximate ones for the higher-$l$ states are known for this potential \cite{hamzavi2012approximate,Asadi:2016ybp}. Thus, one can approximate the binding energies for the $\ket{\chi\chi^*}$ states -- relative to the sum of the free-particle masses, $2 m_\chi + \delta$ -- by the corresponding Hulth\'{e}n-potential results:
\begin{align}\label{eq:hulthenenergy}
    E_n&=\frac{\kappa_n^2}{m_{\chi}}\\
    \kappa_n&=\frac{1}{2}\bigg(\frac{\alpha_{\text{H}}m_{\chi}-n^2 m_{\text{H}}}{n}\bigg)
\end{align}Here $n$ is the principal quantum number of the bound state. To accurately approximate the Yukawa potential by its Hulth\'{e}n counterpart, a normalization condition has to be imposed upon $m_{H}$, which we shall take to be $m_{H}=\frac{\pi^2}{6}m_{\phi}$, as argued for in \cite{Cassel:2009wt}. With this choice, we can simply replace the coupling $\alpha_H$ by $\alpha$.

The bound states in the $L+S$-even $\ket{\chi\chi}+ \ket{\chi^*\chi^*}$ sector are less amenable to analytic approximation due to the presence of the mass splitting in the potential (Eq.~\ref{potential}). However, their presence is crucial in setting the resonance positions. We will thus seek to study the energy spectrum of these states both analytically and numerically.

\emph{A note on binding energy conventions:} Hereafter we will always state binding energies $E$ relative to $2 m_\chi$, in order to have a common mass scale, even when the bound state has $\chi^*$ constituents. We will also quote the binding energies as positive values. In other words, we choose $E$ so that the mass of the bound state is $2 m_\chi - E$. Under this convention, Eq.~\ref{eq:hulthenenergy} gives an estimate for the binding energies of $L+S$-odd states as:
\begin{align}\label{eq:oddjstates}
    E&\approx \frac{\alpha^2 m_\chi}{4 n^2} \bigg(1 - \frac{(\pi^2/6) n^2 m_\phi}{\alpha m_\chi}\bigg)^2 - \delta
\end{align}

We can define a dimensionless binding-energy parameter $\epsilon_E \equiv E/(\alpha^2 m_\chi)$; the approximate $L+S$-odd bound-state spectrum then becomes:
\begin{align}\label{eq:oddjstatesnodim}
    \epsilon_E &\approx \frac{1}{4 n^2} \bigg(1 - \frac{\pi^2}{6} n^2 \epsilon_\phi\bigg)^2 - \frac{\epsilon_\delta^2}{2}
\end{align}

\subsection{Numerical calculation of the $L+S$-even spectrum}

We are interested in solving the eigenvalue problem \begin{equation}\label{eq:eigenvalue}
    \hat{H}\ket{\psi}=-E \ket{\psi}
\end{equation}where $E > 0$ is the binding energy (as defined above) and $\hat{H}$ is the Hamiltonian corresponding to the potential in Eq.~\ref{potential}. Inserting a complete set of states $\ket{j}$, we obtain that \begin{equation}
    H_{ij}c_j=-E c_{i}
\end{equation}where $H_{ij} = \langle i | H | j \rangle$ and $c_{i}=\braket{i|\psi}$. A complete set of states would exactly solve this problem, but we can determine the eigenvalues $E_n$ to any desired accuracy by using a sufficiently large finite number of states $|i\rangle$ \cite{Asadi:2016ybp}.

Following \cite{Asadi:2016ybp}, we use the bound-state wavefunctions of the Coulomb potential with strength $\alpha$ as our basis set; that is, the potential for the basis states is $\frac{1}{r}$ in the scaled coordinates.  This is motivated from the fact that in the limit $\epsilon_\delta=\epsilon_\phi=0$, the matrix potential decouples into an attractive and repulsive Coulomb potential. More explicitly, the requisite basis is constituted by $\begin{bmatrix} \psi_{nlm}\\
0\end{bmatrix}$ and  $\begin{bmatrix} 0\\
\psi_{nlm}\end{bmatrix}$, where $\psi_{nlm}$ is the Couloumbic bound-state wavefunction characterized by the quantum numbers $n,l,m$. We simplify our work further by observing that for a bound state characterized by angular momentum $l$, it suffices to use only the Coulomb bound states having the same quantum number $l$. Note that throughout, we have also fixed $m=0$, since it is easily seen that the binding energies will be degenerate in quantum number $m$. In order to get convergence on the eigenvalues, we used 30 such states. As a numerical check, we ensured that the Couloumbic binding energies were recovered in the limit $\epsilon_{\phi}=\epsilon_{\delta}=0$. As a further check, we ensured convergence by examining that the binding energies were similar when the number of basis states was changed to 40, 50 and 60 respectively.

\begin{figure}[tbp]
    \includegraphics[width=\textwidth]{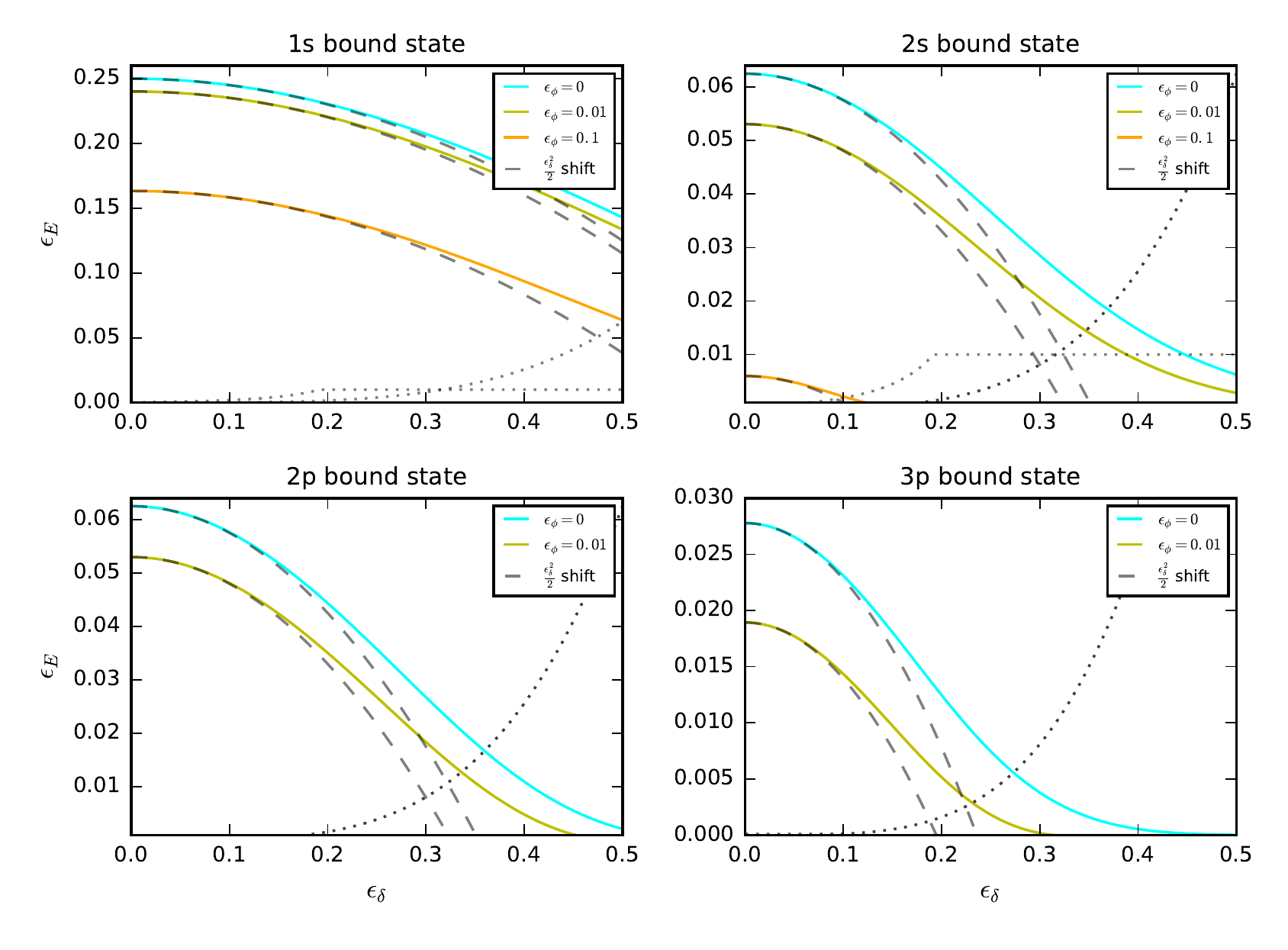}
    \caption{Dimensionless binding energy $\epsilon_E \equiv E(\epsilon_\phi,\epsilon_\delta)/m_{\chi}\alpha^2$ vs $\epsilon_{\delta}$ for the 1$s$ (\emph{top left}), 2$s$ (\emph{top right}), 2$p$ (\emph{bottom left}), and 3$p$ (\emph{bottom right}) states at fixed $\epsilon_\phi$. Blue,green, and orange lines correspond to $\epsilon_{\phi}= 0, 0.01, 0.1$ respectively. The dashed gray lines show the corresponding analytic estimates $(\epsilon_{\delta},\widetilde{E}_b(\epsilon_\phi,0)-\epsilon_{\delta}^2/2)$, where $\widetilde{E}_b$ is the dimensionless binding energy. The curves corresponding to a shift in energy of $\epsilon_{\delta}^2/2$ match  their numerical counterparts well for small $\epsilon_{\delta}$.  Dotted lines indicate the function $\epsilon_\delta^4$ in all four panels; in the upper two panels, we also plot $\text{min}(\epsilon_\phi^2/\ln(\epsilon_\phi/\epsilon_\delta^2)^2,\ \epsilon_{\phi}^2)$ for $\epsilon_\phi=0.1$ (for smaller $\epsilon_\phi$ this constraint is not relevant for any of the bound states we examine). The expected region of validity for the analytic estimates is above and to the left of these curves.
    \label{fig:binding energy}}
\end{figure}

\subsection{Analytic estimates for the $L+S$-even spectrum}

As discussed previously, the bound-state wavefunction should largely have support at small $r$, where the potential is large compared to the mass splitting. In this small-$r$ regime, the mixing between the eigenstates is suppressed, as discussed in Section \ref{sec:general considerations}. Accordingly, we can associate the bound state entirely with the eigenstate that experiences an attractive potential, and read off the potential for that eigenstate from the eigenvalue $\lambda_-$ in Eq.~\ref{smallr}, with the replacement of $-\epsilon_v^2$ with the dimensionless binding-energy parameter $\epsilon_E$ ($\equiv E/\alpha^2 m_\chi$):
\begin{equation}  \lambda_- \approx \epsilon_E +\frac{l(l+1)}{r^2}+\frac{\epsilon_{\delta}^2}{2}- \frac{e^{-\epsilon_{\phi}r}}{r}+\mathcal{O}(r).\end{equation}

We see that to lowest order, the effect of switching on the mass splitting in this case is to simply shift the binding energy parameter $\epsilon_E$ by $\epsilon_\delta^2/2$. Explicitly, suppose the Yukawa potential with the same $\epsilon_\phi$ but no mass splitting has a spectrum of bound states with energies $\epsilon_E^\prime$. Then the spectrum of bound states in the case with a mass splitting will (to the degree that this approximation is valid) satisfy $\epsilon_E + \epsilon_\delta^2/2 = \epsilon_E^\prime$.

In terms of the physical bound-state energies, we can write this result as:
\begin{equation}\label{eq:approx}
    E(\epsilon_{\phi},\epsilon_{\delta})=E(\epsilon_{\phi},0)-\delta.
\end{equation}

We expect this approximation to break down once the radius of the bound state becomes comparable to the crossover radius where the mass splitting term is comparable to the Yukawa potential. We can estimate the typical momentum of a particle in the bound state as $p\sim\sqrt{m_\chi E}$; rescaling the radial coordinate as previously, we obtain an estimate for the dimensionless radius of the bound state:
\begin{equation} r_B \equiv \alpha m_\chi (m_\chi E)^{-1/2} = (\alpha^2 m_\chi/E)^{1/2} = 1/\sqrt{\epsilon_E}.\end{equation}

The crossover radius $r_C$, in the dimensionless rescaled units, is defined by $\epsilon_\delta^2 = e^{-\epsilon_\phi r_C}/r_C$. If $\epsilon_\delta^2 \gg \epsilon_\phi$, then we have $r_C \approx 1/\epsilon_\delta^2$, as the exponent in the Yukawa potential is negligible at $r=r_C$ in this case. The validity condition $r_B \lesssim r_C$ translates in this case to $\epsilon_E \gtrsim \epsilon_\delta^4$. In the opposite case, where $\epsilon_\delta^2 \ll \epsilon_\phi$, the crossover will be induced by the exponential suppression, and we expect $r_C\sim 1/\epsilon_\phi$. We can get a somewhat better estimate by substituting $r_C\approx 1/\epsilon_\phi$ where it appears \emph{outside} the exponent, in the defining equation $\epsilon_\delta^2 = e^{-\epsilon_\phi r_C}/r_C$. Thus we obtain $r_C \approx (1/\epsilon_\phi) \ln (\epsilon_\phi/\epsilon_\delta^2)$. Requiring $r_B \lesssim r_C$ then demands that $\epsilon_E \gtrsim \epsilon_\phi^2/\ln(\epsilon_\phi/\epsilon_\delta^2)^2$, in the regime where $\epsilon_\phi \gg \epsilon_\delta^2$. 

Thus in general this approximation should be valid when:
\begin{equation}\label{eq:validity} \epsilon_E = \frac{E}{\alpha^2 m_\chi} \gg \begin{cases} \epsilon_\phi^2/\ln(\epsilon_\phi/\epsilon_\delta^2)^2, & \epsilon_\delta^2 \ll \epsilon_\phi, \\
\epsilon_\delta^4 & \epsilon_\delta^2 \gg \epsilon_\phi.\end{cases}
\end{equation}
For the intermediate region where $\epsilon_\delta^2 \sim \epsilon_\phi$, both limits become $\epsilon_E \gg \epsilon_\phi^2$. Thus we can estimate the validity constraint over the full region as:
\begin{equation}\epsilon_E  \gg \text{max}\left(\epsilon_\delta^4,\text{min}\left[\epsilon_\phi^2,\epsilon_\phi^2/\ln(\epsilon_\phi/\epsilon_\delta^2)^2\right]\right).
\end{equation}
We have confirmed numerically that this is a reasonable approximation (within a few tens of percent) to $\epsilon_E \gg 1/r_C^2$.

We show the results of the analytic approximation, together with the numerically computed binding energies, in Fig.~\ref{fig:binding energy}. We display results for the $1s, 2s, 2p$ and $3p$ states, as a function of $\epsilon_{\delta}$ at several fixed values of $\epsilon_{\phi}$. To employ Eq.~\ref{eq:approx}, we compute $E(\epsilon_\phi,0)$ numerically in each case (this term can also be estimated analytically using results for the Hulth\'{e}n potential, as we will discuss below). We see that indeed the approximation quite accurately captures the shift in the bound state energies, with the expected breakdown of the approximation once the binding energies become sufficiently small. We overplot dotted lines corresponding to the validity criteria $\epsilon_E \gtrsim \epsilon_\delta^4$ and $\epsilon_E \gtrsim \text{min}\left[\epsilon_\phi^2,\epsilon_\phi^2/\ln(\epsilon_\phi/\epsilon_\delta^2)^2\right]$; the approximation is expected to be valid in the region well above both lines. These estimates seem adequate to characterize roughly where there is a $\mathcal{O}(1)$ divergence between the true and predicted bound-state energies.

For sufficiently shallow bound states, where this approximation will eventually break down, it may be possible to make further progress using the universal characterization of the near-threshold bound-state properties in terms of the scattering length (which approaches infinity for zero-energy bound states) \cite{Braaten:2013tza,Laha:2013gva}. We leave this to future work.

\subsection{Deriving the energy shifts with first-order perturbation theory}

The constant offset in the bound-state energies due to the mass splitting (e.g. Eq.~\ref{eq:approx}) can also be obtained from first-order perturbation theory; this approach makes it easy to see how this result generalizes beyond the pseudo-Dirac case. Suppose we can write the non-relativistic Hamiltonian in the form \begin{equation}
    H=H_0+\Delta H
\end{equation}where $H_0$ is the Hamiltonian in the zero mass-splitting case, and $\Delta H$ is the constant offset matrix induced by mass splitting. In the pseudo-Dirac case,
\begin{equation} \Delta H = 2 \delta \begin{pmatrix} 0 & 0 \\ 0 & 1 \end{pmatrix}.\end{equation}

If we treat $\Delta H$ as a perturbation, then using first-order perturbation theory we find that the correction to the bound-state energy is given by \begin{align}
    \Delta E&=\bra{\psi}\Delta H\ket{\psi}
\end{align}where $\ket{\psi}=\begin{pmatrix}\psi_1(r)\\ \psi_2(r)\end{pmatrix}$ is the bound-state wavefunction in the absence of the mass splitting. 

If it is the case that (1) $H_0$ admits approximately $r$-independent eigenvectors $\eta$, so that we can write $\ket{\psi}=\psi(r)\eta$ where $\psi(r)$ is a canonically normalized scalar wavefunction and $\eta$ is a $r$-independent vector with $\eta^\dagger \eta = 1$, and (2) $\Delta H$ is independent of $r$, fixed solely by the mass splitting, then we can write:
\begin{equation} \Delta E = \eta^\dagger \Delta H \eta \int d^3 r \psi^*(r) \psi(r) = \eta^\dagger \Delta H \eta.\end{equation}

Thus the first-order shift is simply determined by the degree to which the bound state overlaps with the mass splitting matrix. If the bound state is completely constituted of one of the two-body states in the spectrum, its energy is offset by exactly the mass splitting of that state from $2 m_\chi$, as one would expect. Where there are only two possible two-particle states (for even $L+S$) and the mass splitting is $2\delta$, the first-order shift is given simply by $|C|^2 \times 2 \delta$, where $C$ is the second component of $\eta$.

For the dark $U(1)$ (pseudo-Dirac) case with even $L+S$, \begin{align}
    H_0&=-\nabla^2-\frac{e^{-\epsilon_{\phi}r}}{r}\sigma_x,& \eta&=\frac{1}{\sqrt{2}}\begin{pmatrix}1\\1\end{pmatrix}
\end{align}giving us that $C=1/2$ and the binding energy is given approximately by:\begin{equation}
    E=E_0-\delta
\end{equation}which we previously obtained directly from the eigenvalues (Eq.~\ref{eq:approx}).

This alternative derivation has the advantage of being easily applicable to general Hamiltonians of similar form; it manifestly demonstrates that the prefactor in the constant energy offset is determined purely by the fraction of the bound state in the heavier two-particle states of the system.

This approach also clarifies the expected range of validity of the approximation; in the pseudo-Dirac case, it is valid up to $\mathcal{O}(\delta^2) \sim \mathcal{O}(\epsilon_\delta^4)$ terms, and so we expect the corrections to the binding energy from these next-order terms to become large when $\epsilon_E \sim \epsilon_\delta^4$, as previously discussed. This argument further suggests that in the pseudo-Dirac case the constant-energy-shift approximation may still be valid regardless of $\epsilon_\phi$, provided only that $\epsilon_E \gg \epsilon_\delta^4$, since for this $H_0$ the eigenvectors are $r$-independent for arbitrary $\epsilon_\phi$ so long as $\epsilon_\delta = 0$.

\subsection{Shifts to the resonance positions}

\begin{figure}[h]
    \includegraphics[width=\textwidth]{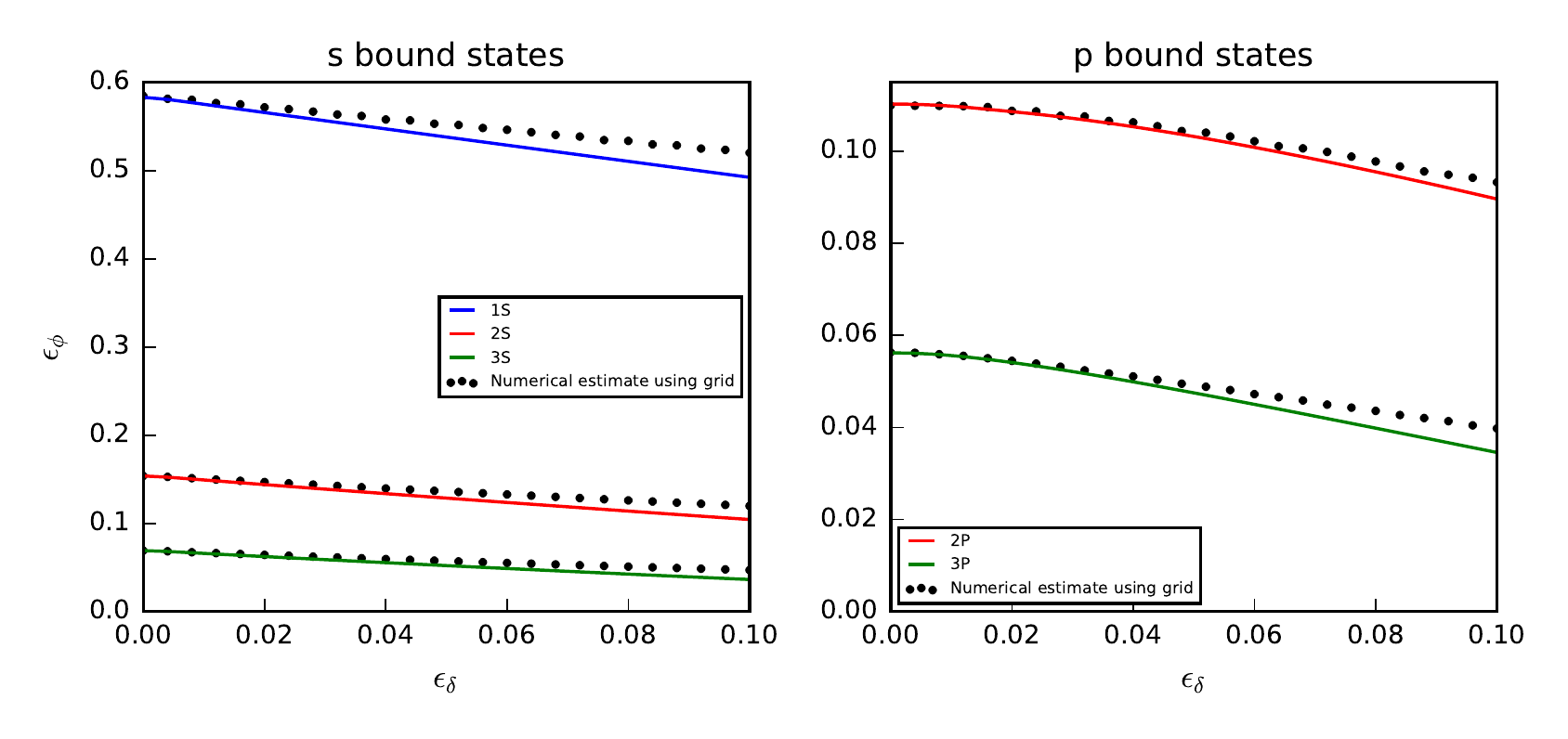}
    \caption{Resonance positions for the $1s,2s,3s$ singlet states (\emph{left panel}) and $2p, 3p$ triplet states (\emph{right panel}). The plots show the value of the dimensionless $\epsilon_{\phi}$ parameter at which the bound state energy becomes zero, as a function of $\epsilon_\delta$. The solid lines indicate the solution for $\epsilon_\phi$ obtained by solving Eq.~\ref{eq:approx} for $E=0$, while the dots indicate numerical calculation of the resonance positions from directly evaluating the bound state energies as a function of $\epsilon_{\phi}$ and $\epsilon_\delta$.
    }
    \label{fig:respos}
\end{figure}

The simple behavior of the bound-state energies in the presence of the mass splitting gives us some analytic control over the positions of the resonances in the Sommerfeld enhancement, corresponding to zero-energy bound states, although (from the arguments in the previous subsection) we expect our approximation to eventually fail in the zero-binding-energy limit. We can test this by computing the values of $\epsilon_\phi$ for which $\epsilon_E(\epsilon_\phi,\epsilon_\delta)=0$, calculated fully numerically as a function of $\epsilon_\delta$, with the values obtained by solving for $\epsilon_E(\epsilon_\phi,\epsilon_\delta)=0$ using Eq.~\ref{eq:approx}. In the latter case, we determine $\epsilon_E(\epsilon_\phi,0)$ numerically as a function of $\epsilon_\phi$, and solve for $\epsilon_E(\epsilon_\phi,0) = \epsilon_\delta^2/2$ as a function of $\epsilon_\delta$. For the bound states $1s,2s,2p,3s,3p$, we find reasonable agreement of the resonance positions with the semi-analytic prediction, at least for small $\epsilon_{\delta}$, as is shown in Fig.~\ref{fig:respos}.

From Fig.~\ref{fig:respos} we can also infer that the shift in the resonance value of $\epsilon_\phi$ is roughly linear in $\epsilon_\delta$. This behavior was already found for $s$-wave resonances in Ref.~\cite{slatyer2010sommerfeld}. We can see why this occurs analytically from the estimate for the binding energies of the Hulth\'{e}n potential given in Eq.~\ref{eq:hulthenenergy}. Choosing $\alpha_H=\alpha$ as previously, and writing $\epsilon_{H}=m_H/\alpha m_{\chi}$, we can write the Hulth\'{e}n spectrum in the form:
\begin{equation} \epsilon_E(\epsilon_H,0) \approx 
    \frac{\left(1 - n^2 \epsilon_H \right)^2}{4 n^2} \end{equation}

Suppose we approximate the Yukawa potential (with no mass splittings) by a Hulth\'{e}n potential with $\epsilon_H = q \epsilon_\phi$ for some constant $q$ (usually taken to be $q=\pi^2/6$), so $\epsilon_E(\epsilon_\phi,0) \approx \left(1 - n^2 q \epsilon_\phi \right)^2/4 n^2$. Then using Eq.~\ref{eq:approx}, the resonance condition $\epsilon_E(\epsilon_\phi,\epsilon_\delta)=0$ becomes:
\begin{equation}\label{eq:eh}
\epsilon_\phi = \frac{1}{n^2 q} \left(1 - \sqrt{2} n \epsilon_\delta \right)
\end{equation}
This yields the expected linear scaling with $\epsilon_\delta$. The slope of the scaling relation depends on the normalization factor $q$ in a trivial way; $q=\pi^2/6$ works well for the $l=0$ case studied in Ref.~\cite{slatyer2010sommerfeld}, but for higher partial waves, we find that slightly modified values of $q$ may be preferred. (Furthermore, for higher partial waves Eq.~\ref{eq:hulthenenergy} is not exact, even for the Hulth\'{e}n potential.) Nonetheless, this argument is sufficient to demonstrate that the approximately linear shift in the resonance positions with increasing $\epsilon_\delta$ is expected to hold true for all $l$, and to a first approximation depends only on $\epsilon_\delta$ and the principal quantum number of the relevant bound state.

\section{Application to electroweak bound states}
\label{sec:wino}

The basic principles laid out in the previous section are not specific to the model we consider. As another example where they may be applicable, consider the case of $SU(2)_{L}$ triplet wino dark matter. The DM is a Majorana fermion, denoted $\chi^0$, and the lightest member of the multiplet; the rest of the multiplet forms a charged Dirac fermion which we will denote $\chi^-$, the chargino. The gauge group is $SU(2)_{L}\times U(1)_{Y}$, and interactions involving the DM and the chargino are mediated by the massive gauge bosons $W^{\pm}$ and $Z$, as well as the photon. The presence of charged gauge bosons allows capture into bound states via a new channel wherein $W^{\pm}$ could itself emit a photon \cite{Asadi:2016ybp}.

As in the simple pseudo-Dirac model, the relevant potentials differ for $L+S$-even and $L+S$-odd states. The neutral bound states supported by the latter potential are again relatively simple, since the two-state system with $L+S$-odd has no $\chi^0\chi^0$ component, and so consists only of $\chi^+ \chi^-$ bound states supported by $Z$ and $\gamma$ exchange (as studied in e.g. Ref.~\cite{Zhang:2013qza}). The DM-chargino mass splitting thus plays no interesting role for this sector.

Consequently, we focus again on the case of even $L+S$, where the potential \cite{Asadi:2016ybp} is given by \begin{align}
    V_{\text{even }L+S}(r)=\begin{bmatrix}
    0 & -\sqrt{2}\alpha_{W}\frac{e^{-m_{W}r}}{r}\\
    -\sqrt{2}\alpha_{W}\frac{e^{-m_{W}r}}{r}& 2\delta-\frac{\alpha}{r}-\frac{\alpha_{W}\cos^2{\theta_{w}}e^{-m_{Z}r}}{r}
    \end{bmatrix}
\end{align}

Let us scale the coordinate $r$ by $\frac{\hbar}{\alpha_W m_{\chi}c}$, yielding the dimensionless analog of Eq.~\ref{matrix} for the wino case, which is \begin{equation}
    V_{\text{even }L+S}(r)=\begin{bmatrix}
    0 & -\sqrt{2}\frac{e^{-\epsilon_{W}r}}{r}\\
    -\sqrt{2}\frac{e^{-\epsilon_{W}r}}{r}& \epsilon_{\delta}^2-\frac{\sin^2\theta_w}{r}-\frac{\cos^2{\theta_{w}}e^{-\epsilon_{Z}r}}{r}
    \end{bmatrix}
\end{equation}where we have defined \begin{align}
    \epsilon_{W}&=\frac{m_{W}}{m_{\chi}\alpha_W}& 
    \epsilon_{Z}&=\frac{m_{Z}}{m_{\chi}\alpha_W}&
    \epsilon_{\delta}&=\sqrt{\frac{2\delta}{m_{\chi}\alpha_W^2}}
\end{align}
Note that the binding energies are now expressed as multiples of $m_{\chi}\alpha_W^2$, $\epsilon_E = E/(\alpha_W^2 m_\chi)$. We can again evaluate the eigenvalues of the potential matrix:
\begin{align}\label{eweigenvalues}
    \lambda_{\pm}(r)=\frac{1}{2}\bigg(\epsilon_{\delta}^2-\frac{\sin^2\theta_w}{r}-\frac{\cos^2{\theta_{w}}e^{-\epsilon_{Z}r}}{r}\bigg)\pm \sqrt{\frac{1}{4}\bigg(\epsilon_{\delta}^2-\frac{\sin^2\theta_w}{r}-\frac{\cos^2{\theta_{w}}e^{-\epsilon_{Z}r}}{r}\bigg)^2+\bigg(\sqrt{2}\frac{e^{-\epsilon_{W}r}}{r}\bigg)^2}
\end{align} 

Again, we will be interested in the behavior of the potential over the support of the bound state, i.e. $r \lesssim 1/\sqrt{\epsilon_E}$. In the regime where $\epsilon_E \gg \epsilon_W^2, \epsilon_Z^2,\epsilon_\delta^4$, then within this radius, the Yukawa terms may be replaced by Coulombic terms, and furthermore these Coulombic terms dominate over the $\epsilon_\delta^2$ term. In this case, the eigenvalues become:
\begin{align}
    \lambda_{\pm}(r)\approx&\frac{1}{2}\bigg(\epsilon_{\delta}^2-\frac{\sin^2\theta_w}{r}-\frac{\cos^2{\theta_{w}}}{r}\bigg)\pm \sqrt{\frac{1}{4}\bigg(\epsilon_{\delta}^2-\frac{\sin^2\theta_w}{r}-\frac{\cos^2{\theta_{w}}}{r}\bigg)^2+\bigg(\frac{\sqrt{2}}{r}\bigg)^2}\nonumber\\
    \approx&\frac{1}{2}\bigg(\epsilon_{\delta}^2-\frac{1}{r}\bigg)\pm\sqrt{\frac{1}{4}\bigg(\epsilon_{\delta}^2-\frac{1}{r}\bigg)^2+\frac{2}{r^2}}\nonumber\\
    \approx&\frac{1}{2}\bigg(\epsilon_{\delta}^2-\frac{1}{r}\bigg)\pm\frac{3}{2r}\sqrt{1-\frac{2\epsilon_{\delta}^2 r}{9}}\nonumber\\
    \approx&\frac{1}{2}\bigg(\epsilon_{\delta}^2-\frac{1}{r}\bigg)\pm\bigg(\frac{3}{2r}-\frac{\epsilon_{\delta}^2}{6}\bigg)\nonumber
\end{align}

Thus we find a pair of attractive and repulsive potentials, with the attractive one being stronger, with offsets due to the mass splitting similar to what we observed in our toy model:
\begin{align}
    \lambda_{-}(r)\approx-\bigg(\frac{2}{r}-\frac{2\epsilon_{\delta}^2}{3}\bigg)& &\lambda_{+}(r)\approx\frac{1}{r}+\frac{\epsilon_{\delta}^2}{3}
\end{align}
In the limit of zero mass splitting (or small $r$), these potentials correspond to full restoration of the $SU(2)_L$ symmetry \cite{Asadi:2016ybp}. Since the repulsive potential cannot accommodate bound states, the effect of the mass splitting is simply to shift the bound state energies supported by the attractive potential to $\epsilon_E(\epsilon_\delta) \approx \epsilon_E(\epsilon_\delta=0) -(2/3) \epsilon_\delta^2$, or in terms of the energies without rescaling,
\begin{equation} E(\epsilon_\delta) \approx E(\epsilon_\delta=0) - (4/3) \delta.\label{eq:winoapprox}\end{equation}

We can also apply the perturbation-theory approach here, in the limit where $\epsilon_W$, $\epsilon_Z$ approach zero and so the potential can be diagonalized in a $r$-independent way. As in the pseudo-Dirac case, we have $\Delta H = \begin{pmatrix} 0 & 0 \\ 0 & 2\delta\end{pmatrix}$. We can rewrite the Hamiltonian (in rescaled coordinates) in the form:
\begin{align}
     H_0&\approx-\nabla^2-\frac{1}{r}\begin{pmatrix}0&\sqrt{2}\\
     \sqrt{2}&1\end{pmatrix}, & \eta&=\frac{1}{\sqrt{3}}\begin{pmatrix}1\\\sqrt{2}\end{pmatrix}
\end{align}so that the binding energies are approximately given by \begin{equation}
    E=E_0-\frac{2}{3} \times 2 \delta
\end{equation} as in Eq.~\ref{eq:winoapprox}.

\begin{figure}[h]
    \centering
    \includegraphics[scale=1]{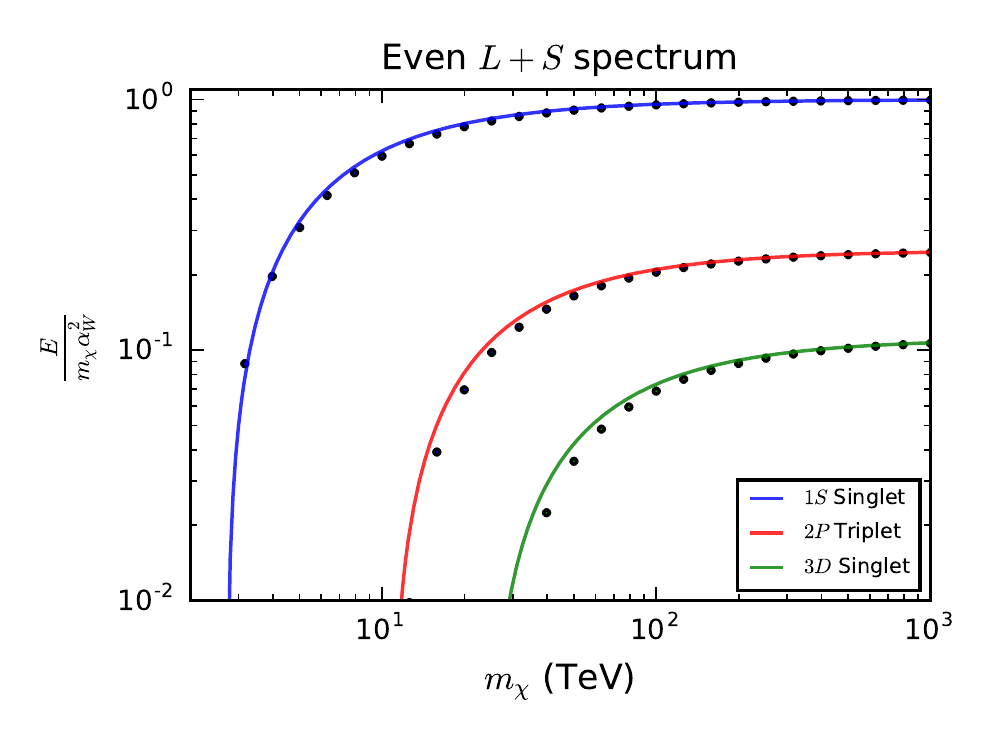}
    \caption{The wino bound-state energy as a function of the wino mass $m_{\chi}$. Here the overplotted points represent numerical evaluations of the bound state energy, while the solid lines are the analytic estimates of Eq.~\ref{eq:winobound}. From top to bottom, the blue, red and green lines describe the $1s$ singlet, $2p$ triplet and $3d$ singlet bound states, respectively.}
    \label{fig:evenls}
\end{figure}

These arguments only hold in full when the potential is essentially Coulombic over the support of the bound states. If the masses of the $W$ and $Z$ bosons are large enough, relative to the DM mass, to significantly perturb the bound state energies, that must also be taken into account. An ad hoc estimate for this effect can be obtained by replacing the Coulombic bound-state energies with those for the Hulth\'{e}n potential, with $m_H \rightarrow (\pi^2/6) m_W$ and $\alpha_H \rightarrow \alpha_W$. Combining this prescription with the shift due to the mass splitting, we obtain the estimate:
\begin{equation}
    E \approx m_{\chi}\alpha_{W}^2\left(\frac{1}{n}-\frac{n\ m_{W}\pi^2}{12\alpha_W m_{\chi}}\right)^2-\frac{4}{3}\delta, \quad \epsilon_E =\frac{1}{n^2} \left(1-\frac{\pi^2 n^2 \epsilon_W }{12}\right)^2-\frac{2}{3}\epsilon_\delta^2 
    \label{eq:winobound}
\end{equation}
In Fig.~\ref{fig:evenls}, we compare the numerically computed bound-state energies for even $L+S$ to this estimate, and find remarkably good agreement across a broad wino mass range.  Throughout, we fixed $m_W=80.38$ GeV, $m_Z=91.19$ GeV, $\delta=0.17$ GeV, and $\alpha_W=0.0335$. For our numerical computation we use 30 basis states as previously, but with the coupling for the basis wavefunctions set to $2\alpha_W$, as discussed in Ref.~\cite{Asadi:2016ybp}.

\begin{figure}[htb]
    \centering
    \includegraphics[scale=1.1]{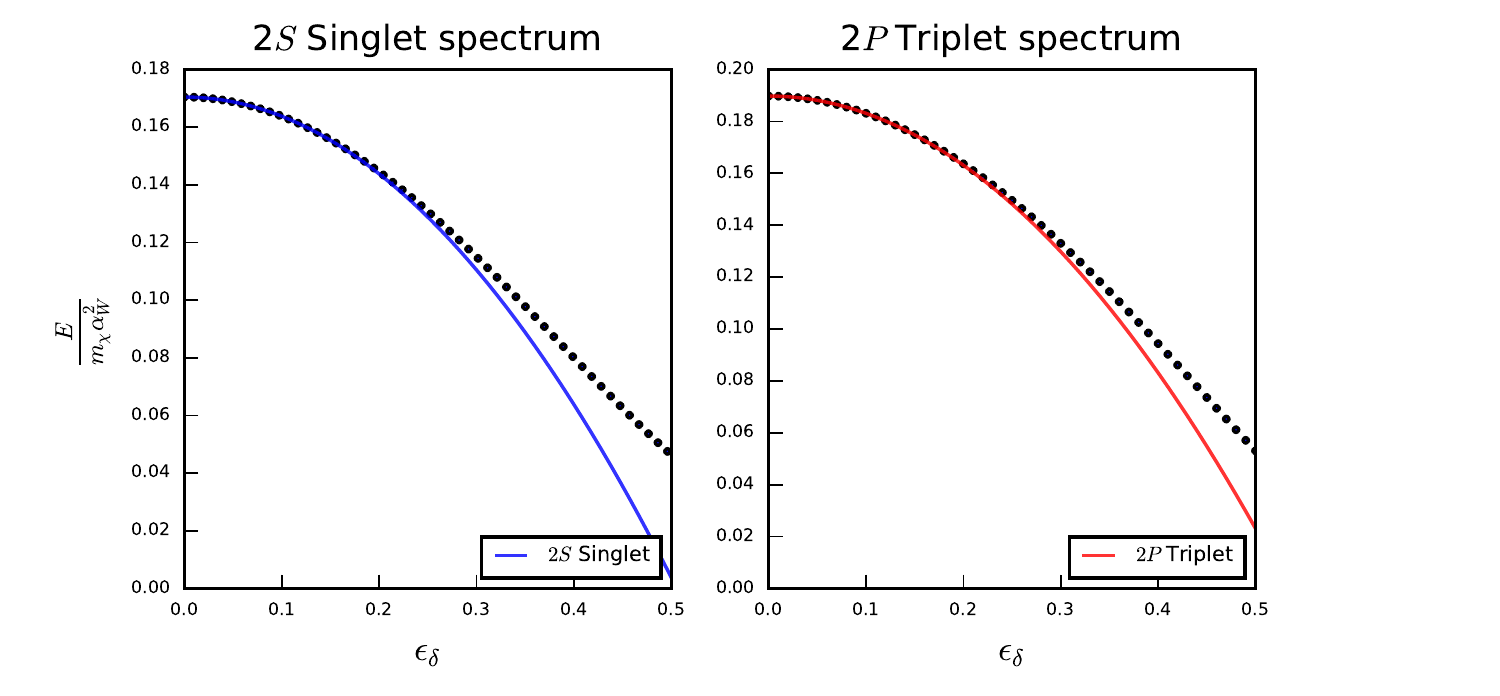}
    \caption{The wino bound-state energy when varying the mass splitting parameter $\epsilon_{\delta}$. The overplotted points represent numerical evaluations of the bound state energy, while the solid lines are the analytic estimates $\epsilon_E(\epsilon_\delta)=\epsilon_E(\epsilon_\delta=0) -(2/3) \epsilon_\delta^2$. In the \emph{left panel} the parameters are $m_{\chi}=50$ TeV, $\epsilon_{W}=0.0479671$; in the \emph{right panel} they are  $m_{\chi}=70$ TeV, $\epsilon_{W}=0.0342622$.}
    \label{fig:winodel}
\end{figure}

As a further check, we examined the behavior of the estimate,
\begin{equation}
    \epsilon_E(\epsilon_\delta) = \epsilon_E(0) -\frac{2}{3} \epsilon_\delta^2
\end{equation}(holding all other parameters fixed) by numerically computing the bound-state energies as a function of the mass splitting $2\delta$, and comparing them to this analytic estimate, for the even-$L+S$ spectrum. We tested the $2s$ singlet and $2p$ triplet cases, and again found reasonable agreement for this simple estimate, as shown in Fig.~\ref{fig:winodel}. Thus, this approach works well even for a more complicated gauge structure such as the wino / $SU(2)$ triplet DM. The approximation of a constant mass-splitting dependent shift to the bound state energies is a generic feature, as demonstrated by the argument from first-order perturbation theory.

\begin{figure}[tb]
    \centering
    \includegraphics[scale=1]{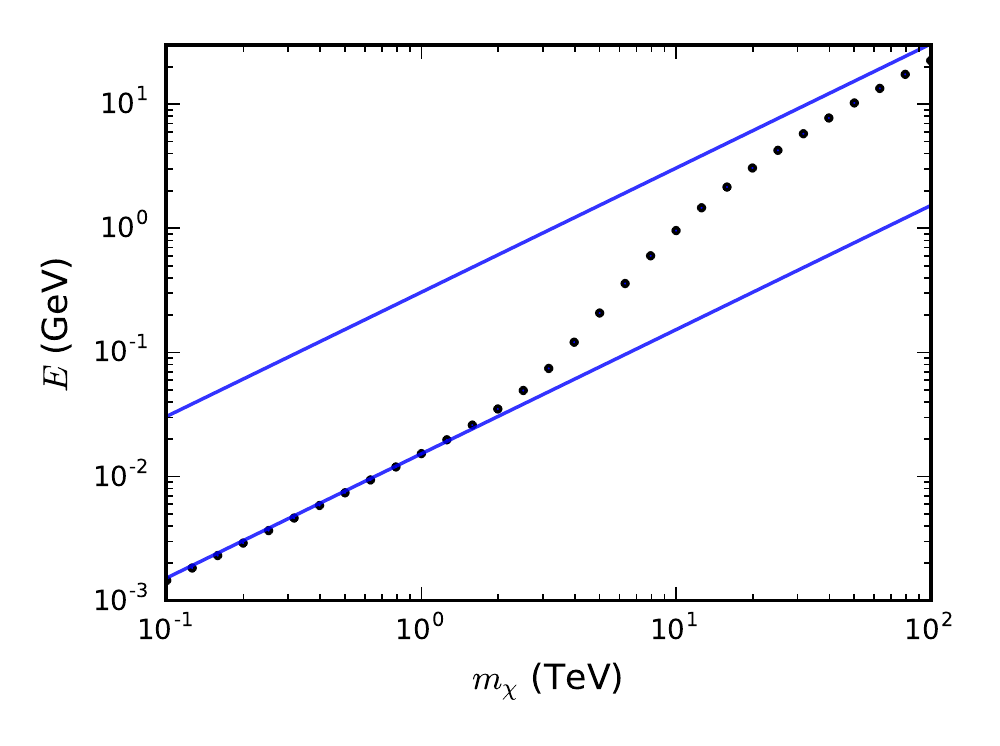}
    \caption{The binding energy for the $1s$ spin-triplet bound state (i.e. with odd $L+S$) for wino DM, relative to $2m_\chi + 2 \delta$ (note the binding energies under our usual convention can thus be obtained by subtracting $2\delta$ from this curve), as a function of the wino mass $m_{\chi}$. Here the black points represent numerical evaluations of the bound state energy, while the solid blue lines are the Coulombic limits with two different couplings, $E=\alpha^2m_{\chi}/4$ and $E=\alpha_W^2m_{\chi}/4$. The effective coupling strength transitions from $\alpha$ at low DM masses (where the $\gamma$ exchange dominates) to $\alpha_W$ at high masses (where $SU(2)$ symmetry is restored).}
    \label{fig:alpha}
\end{figure}

The analysis for the $L+S$-odd case is again simpler; we need only recall that since binding energies are defined relative to $2 m_\chi$, the potential for the odd-$L+S$ case contains a constant offset of $2\delta$ due to the higher mass of its constituents. In the regime where the potential is largely Coulombic, and the mass of the $Z$ boson can be neglected, the binding energies defined relative to the sum of the constituent masses, $2 m_\chi + 2 \delta$, follow the usual Coulomb form $E_n = \alpha_W^2 m_\chi/4 n^2$. Thus under our convention, where the total mass of the state is $2 m_\chi - E$, the binding energies $E$ can be approximated in this case as: \begin{align}
    E =\frac{\alpha_{W}^2m_{\chi}}{4n^2}-2\delta, \quad \epsilon_E = \frac{1}{4 n^2} - \epsilon_\delta^2
\end{align}
In the limit where the $Z$ boson mass is large relative to $\alpha_W m_\chi$, the Coulombic interaction will dominate the potential, and the bound states will have a similar pattern, but with $\alpha_W \rightarrow \alpha$. This transition is demonstrated numerically in Fig.~\ref{fig:alpha}.

\section{The cross section for radiative formation of bound states}
\label{sec:capture}

Having derived and understood the bound state spectrum, we now return to the simple pseudo-Dirac model to explore the impact of turning on the mass splitting on the cross section for radiative bound-state formation. There is a subtlety here: using only the ingredients of our model described thus far, the relevant process is emission of a dark photon with mass $m_\phi$, which in the low-velocity limit requires that the binding energy $E$ exceed $m_\phi$, i.e. $m_\phi$ is parametrically of size $\alpha^2 m_\chi$ or smaller. However, to the degree that we wish to use this model as a testbed for other scenarios with mass splittings between DM states, it is useful to consider the possibility that some lighter (or even massless) particle could couple to the $\chi$ or $\chi^*$ states and be radiated to allow formation of a bound state. For example, a similar behavior occurs naturally in the case of wino or higgsino DM: while the DM itself does not couple to the photon, only to the weak gauge bosons (and possibly the Higgs), it has nearly-degenerate heavier charged partners that play a similar role to the $\chi^*$ in our present scenario, and these partners can emit massless photons to allow for bound-state formation. In this way, the presence of a light particle can be crucial for bound-state formation, even if it does not dominate the potential experienced by the DM particles.

In what follows, in order to build intuition for these expanded cases, we will leave open the possibility that a light vector is emitted, with a mass and coupling to the DM that do not match those of the dark photon that dominates the potential; equivalently, when we study how the squared matrix element for this process is modified by the presence of a mass splitting, we will also show results for regimes where the phase-space factor would vanish if the particle being emitted is a dark photon with mass $m_\phi$. However, we will work under the assumption that the particle emitted is a vector rather than a scalar, and thus the selection rules are the same as they would be for dark photon emission; violating this assumption would modify the matrix element, not just the phase-space factor (see Ref.~\cite{Oncala:2018bvl} for a discussion of bound-state formation through scalar emission).

As in Ref.~\cite{Asadi:2016ybp}, we define the matrix element $\bar M$ for radiative capture in the dipole approximation as:
\begin{equation}\label{matrixelement}
    \bar M\equiv \frac{\varepsilon}{\mu}\cdot\int d^3 r~ \Psi_{\text{scat}}^{\dagger}(r)\ \hat{C}\ \nabla_{r}\Psi_{\text{bound}}(r)
\end{equation} where $\varepsilon$ describes the polarization of the emitted light vector, $\mu = m_\chi/2$ is the reduced mass of the incoming DM particles, and $\hat{C}$ is the matrix that couples the $\ket{\chi\chi^*}$ sector to the $\ket{\chi\chi} + \ket{\chi^* \chi^*}$ sector. If we expand the previous basis to include $\ket{\chi\chi^*}$, writing the state $\alpha \ket{\chi \chi} + \beta \ket{\chi^*\chi^*} + \gamma \ket{\chi \chi^*}$ as $\begin{pmatrix} \alpha \\ \beta \\ \gamma\end{pmatrix}$, then $\hat{C}$ takes the form:
\begin{equation}\hat{C} =\frac{1}{\sqrt{2}} \begin{pmatrix} 0 & 0 & 1 \\ 0 & 0 & 1 \\ 1 & 1 & 0 \end{pmatrix} \end{equation}
The factor of $1/\sqrt{2}$ is needed to account for the fact that the $\ket{\chi \chi^*}$ state could equivalently be described as $\ket{\chi^* \chi}$ \cite{Beneke:2014gja}.

The bound states relevant for the single-dipole-vector capture process (which is expected to dominate the overall rate), lie in the $L+S$-odd $\ket{\chi\chi^*}$ sector, since the initial state is $L+S$-even and the dipole photon radiation gives $\Delta L = \pm 1$, $\Delta S = 0$. 

We solved for the scattering wavefunctions numerically, using the $L+S$-even potential and employing the method of variable phase, as elucidated in detail in Ref.~\cite{Asadi:2016ybp}. This numerical approach was chosen for its superior stability at large $r$, compared to solving the Schr\"{o}dinger equation via brute force using the \texttt{NDSolve} function in \texttt{Mathematica}. For an initial state consisting of two identical fermions,\footnote{Note that as discussed in Ref.~\cite{Asadi:2016ybp}, the presence of identical fermions in the initial state means this cross section is larger than that for non-identical initial-state particles by a factor of 2 when the initial state has $L+S$ even, and is zero for $L+S$ odd.} the matrix element in Eq.~\ref{matrixelement} can be translated to a physical cross section as:
\begin{equation}\label{xsec}
    \sigma v_{\text{rel}}=\frac{\alpha_\text{rad} k}{\pi}\int d\Omega~|\bar{M}|^2.
\end{equation}Here $v_{\text{rel}}$ is the relative velocity between the particles, $\alpha_\text{rad}$ describes the coupling of the emitted light vector to the DM (which may or may not be equal to the coupling $\alpha$ that controls the potential), and the energy of the emitted particle is denoted $k$ and given by:
\begin{align}
k&=E_{\text{scat}}-(2 m_{\chi} - E)\\
&=m_{\chi}\alpha^2(\epsilon_v^2+\epsilon_E)
\end{align}
for a massless or near-massless particle, where $E_\text{scat}$ is the total (mass+kinetic) energy of the initial state, and $E$ is the binding energy of the final-state bound state. (Recall that $\epsilon_E$ contains a contribution of $-\epsilon_\delta^2/2$, as in Eq.~\ref{eq:oddjstatesnodim}, since the bound state is composed of $\chi$ and $\chi^*$.) If the emitted particle has a non-zero mass $m_\text{rad}$, its energy should be modified to:
\begin{align}
    E_k&=\sqrt{k^2+m_\text{rad}^2}\nonumber\\
    &=\frac{E_{\text{scat}}^2+m_\text{rad}^2-(2 m_{\chi}- E)^2}{2E_{\text{scat}}}\nonumber\\
    &\approx \left(E_{\text{scat}}-(2 m_{\chi} - E)\right)\left(1-\frac{m_{\chi}\alpha^2(\epsilon_v^2+\epsilon_E)}{2 E_{\text{scat}}}\right)+\frac{m_\text{rad}^2}{2E_{\text{scat}}}\nonumber\\
    &\approx m_{\chi}\alpha^2\left(\epsilon_v^2+\epsilon_E+\frac{\alpha^2 \epsilon_\text{rad}^2}{4}\right),
\end{align}
where $\epsilon_\text{rad} \equiv m_\text{rad}/(\alpha^2 m_\chi)$. Consequently the emitted particle's momentum $k$ is modified to:
\begin{align} k = \sqrt{E_k^2 - m_\text{rad}^2} \approx m_\chi \alpha^2 \sqrt{(\epsilon_v^2 + \epsilon_E + \alpha^2 \epsilon_\text{rad}^2/4)^2 - \epsilon_\text{rad}^2}.\end{align}

As in \cite{Asadi:2016ybp}, we can simplify Eq.~\ref{xsec} further by writing\begin{equation}
    \bar{M}=A\epsilon\cdot \hat{r}_{m}
\end{equation}for capture into the bound state characterized by the quantum numbers $\{n,l,m\}$, and we perform the angular integral separately. Using considerations of spherical symmetry, Eq.~\ref{xsec} then simplifies to \begin{equation}
    \sigma v_{\text{rel}}= \frac{8\alpha_\text{rad} k|A|^2}{3}
    \label{eq:xsecsimple}
\end{equation} 
Again, this cross section is valid under the assumption that the initial state contains two identical fermions and has $L+S$ even; in a situation where the initial state contains two distinguishable particles, this cross section should be divided by two (and need not be zero for $L+S$ odd). Note this is the cross section for an initial state of fixed spin, rather than a spin-averaged cross section (performing the spin average would introduce another factor of $1/4$ or $3/4$, for $L$-even and $L$-odd initial states respectively).

\begin{figure}[htb]
    \includegraphics[width=\textwidth]{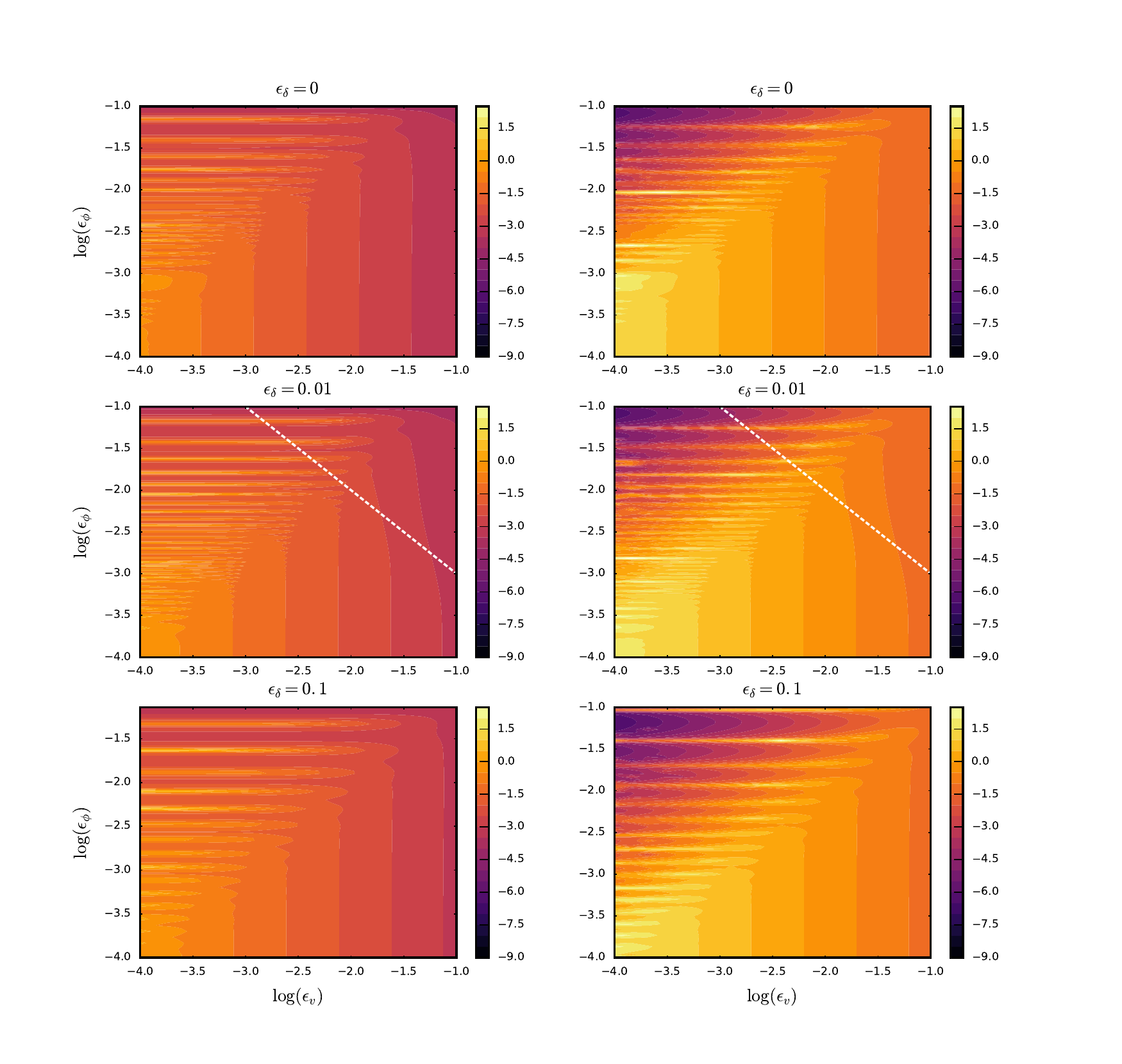}
    \caption{The bound-state formation cross section, plotted as $\log_{10}(\sigma v_\text{rel} m_\chi^2 c /\hbar)$, for $\epsilon_\delta=0$ (\emph{top row}), $\epsilon_\delta=0.01$ (\emph{middle row}) and $\epsilon_\delta=0.1$ (\emph{bottom row}). We show results for capture into the $2p$ spin-singlet bound state from the $s$-wave component of an initial plane wave (\emph{left column}) and for capture into the $1s$ spin-triplet bound state from the $p$-wave component of the initial state (\emph{right column}). The white line indicates the locus of $\epsilon_{v}\epsilon_{\phi}=\epsilon_{\delta}^2$; the cross section in the region below the line is double the cross section at zero mass splitting, at velocities far from resonances (see text for details). 
    \label{fig:contour plots}}
\end{figure}

We numerically evaluated this cross section for a scan over $(\epsilon_v, \epsilon_{\delta}, \epsilon_{\phi})$, using the phase-space factor for a massless vector to examine the maximum amount of parameter space (our results can trivially be rescaled to a different phase-space factor using Eq.~\ref{eq:xsecsimple}). We chose to examine the region where there is a large Sommerfeld enhancement, and the Born approximation for the scattering cross-sections is insufficient, i.e. where the dimensionless parameters are less than unity \cite{slatyer2010sommerfeld}). Contour plots for the dimensionless cross section $\sigma v_{\text{rel}}m_{\chi}^2c/\hbar^2$, broken down by the initial and final quantum numbers, are shown in Fig.~\ref{fig:contour plots}. Throughout, we have set $\alpha_{\text{rad}}=\alpha=0.01$ in our numerical evaluations.

Several broad features emerge from these plots. Let us first comment briefly on the difference between the capture from $s$-wave and $p$-wave initial states, which is manifest even for $\epsilon_\delta=0$. For a $s$-wave initial state, we expect (in the absence of a long-range potential) $\sigma v_\text{rel} \propto v_\text{rel}^0$, while for the $p$-wave contribution we expect $\sigma v_\text{rel} \propto v_\text{rel}^2$. This behavior is observed in the limit where $\epsilon_v \ll \epsilon_\phi$, the upper left corners of the contour plots. However, in the opposite limit where $\epsilon_v \gg \epsilon_\phi$ (lower right corner of the contour plots), we recover Coulombic behavior where the potential is effectively long-range. In this regime, as we will discuss in more detail below, all partial waves contribute with the same $1/v$ scaling. Thus in the $\epsilon_\delta=0$ case we expect (and observe) $\sigma v_\text{rel}$ to rise with decreasing velocity for both $s$- and $p$-wave initial states, before flattening out for the $s$-wave case, and being suppressed at low velocity for the $p$-wave case.

Upon turning on a mass splitting, we find that:
\begin{itemize}
    \item In the case of non-zero $\delta$, in the region of parameter space where $\epsilon_{v}\epsilon_{\phi}\lesssim \epsilon_{\delta}^2$, the cross section is doubled relative to the case with zero mass splitting. This is most clearly illustrated by comparing the first and second rows of  Fig.~\ref{fig:contour plots}, and in the second row, observing the cross section at fixed $\epsilon_{v}$ as $\epsilon_{\phi}$ is varied across the boundary $\epsilon_\phi \sim \epsilon_\delta^2/\epsilon_v$. 
    
    This behaviour has been noted previously in the context of the Sommerfeld enhancement, and can be explained by a transition between ``adiabatic'' and ``non-adiabatic'' regimes \cite{schutz2015self,zhang2017self}, with regard to the rotation of the eigenvectors with radial distance from the origin. The argument is that if the diagonalizing matrix varies sufficiently slowly with respect to $r$, then a large-$r$ asymptotic state consisting purely of $\ket{\chi\chi}$ (the lower-energy state at large $r$) will smoothly transition to a small-$r$ state that is dominated by the eigenvector experiencing an attractive potential (the lower-energy state at small $r$). In contrast, in the case with $\epsilon_\delta=0$, there is no evolution of the diagonalizing matrix with $r$, and the small-$r$ wavefunction has the same contributions from the repulsed and attracted eigenvectors as the large-$r$ state -- in this model, those contributions are equal. Since only the attracted eigenvector yields a significant contribution to bound-state formation, the probability of bound-state formation is reduced by a factor of 2 in the Coulombic case relative to the adiabatic case.

    The adiabatic regime is characterized by the criterion \cite{schutz2015self}:  \begin{equation}\label{eq:adiabatic_criterion}
        \epsilon_{v}\epsilon_{\phi}\leq\epsilon_{\delta}^2
    \end{equation}
    
    In the wino case, we expect similar behavior in the same regime, except that the cross section will be enhanced by a factor of 3 (rather than 2) relative to the case with no mass splitting. The argument is almost identical to that of the dark $U(1)$ case, and the factor is 3 in this case since the large-$r$ asymptotic state has only a $1/3$ overlap with the attracted eigenvector \cite{Asadi:2016ybp}. This factor-of-3 enhancement is observed in the numerical results of Ref.~\cite{Asadi:2016ybp}. 

    \item As the mass splitting turns on, the resonances in the cross section for capture are enhanced, and the resonance positions undergo a shift relative to the Coulombic case, according to Eq.~\ref{eq:eh}, when the kinetic energy becomes negligible compared to the mass splitting, as illustrated in Fig.~\ref{fig:adiaplot}. This behavior is due to a modification to the initial-state wavefunction, with similar effects being observed in studies of the Sommerfeld enhancement \cite{slatyer2010sommerfeld}; thus the resonance positions are determined by bound states in the $L+S$-even sector, even though the bound state being formed in this case has $L+S$ odd. Because the resonance positions now also depend on the importance of the mass splitting term compared to the kinetic energy, it is possible for a particular choice of $(\epsilon_\phi, \epsilon_\delta)$ to lie on a resonance at one velocity but not at a lower velocity, leading to non-monotonicity in the capture rate even for a $s$-wave initial state (similar behavior is again observed in the multi-state Sommerfeld case \cite{slatyer2010sommerfeld}). This behavior is demonstrated in the right panel of Fig.~\ref{fig:adiaplot}.
    
    \begin{figure}[htb]
        \centering
        \includegraphics[scale=1]{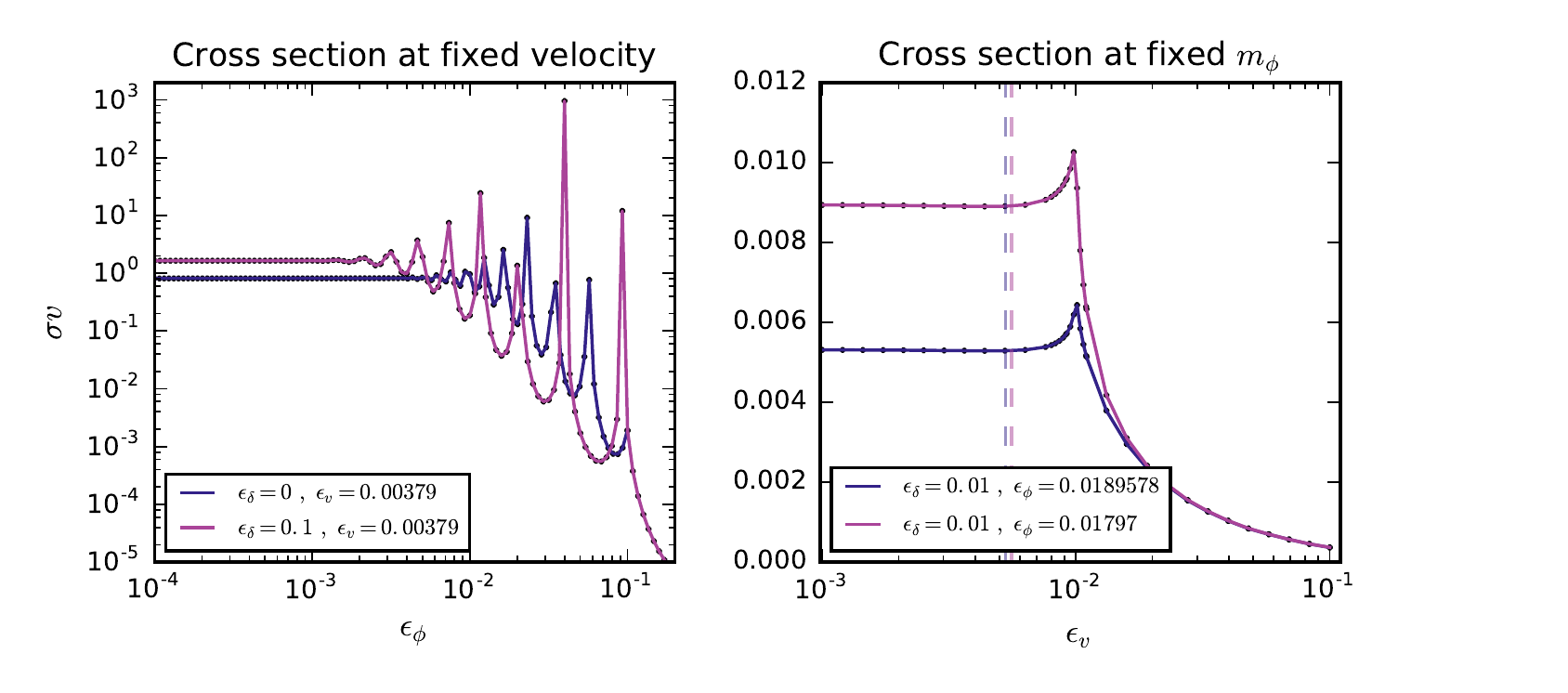}
        \caption{Radiative capture cross section $\sigma v$ at fixed $\epsilon_{v}$ (\emph{left panel}) and fixed $\epsilon_{\phi}$ (\emph{right panel}) for an initial $s$-wave state and $2p$ bound state, taking $\alpha_\text{rad}=0.01$ and assuming emission of a massless vector (these results can be rescaled to a massive vector by insertion of a modified phase space factor). In the right panel, colored dashed vertical lines indicate the locus of $\epsilon_v=\epsilon_{\delta}^2/\epsilon_{\phi}$, with the region on their left corresponding to the adiabatic regime. We observe the enhancement of the cross section for non-zero $\delta$ in the adiabatic regime, in the right panel, and the $\delta$-dependent shift in the resonance positions, in the left panel.}
        \label{fig:adiaplot}
    \end{figure}

 \end{itemize}
 
 The two generic behaviors described above (adiabatic enhancement and shifting of the resonance structure) originate from the properties of the initial-state wavefunction, and are thus also seen in the calculation of the Sommerfeld enhancement to short-distance annihilation processes (which is set by the initial-state wavefunction evaluated at the origin).
    
    The impact of changing the initial-state wavefunction on the matrix element for bound-state formation is in principle not identical to its effect on the matrix element for short-distance processes such as annihilation; the former involves a non-trivial integration with respect to $r$, as the bound-state formation process is localized within the region $r \lesssim 1/(\alpha m_\chi)$, rather than at the origin. However, if the scale $1/(\alpha m_\chi)$ is small compared to the scales over which the initial-state wavefunction varies significantly, we might expect the fact that the interaction is not zero-range to have only a mild effect. The unperturbed initial-state wavefunction has a natural scale of $1/(m_\chi v)$, the de Broglie wavelength of the incoming particles, and so we might expect the bound-state formation process to be effectively short-range when $m_\chi v \ll m_\chi \alpha$, i.e. $\epsilon_v \ll 1$ (this argument was also made in Ref.~\cite{Finkbeiner:2010sm}). However, it is not clear that this argument will still hold when the initial-state wavefunction is significantly deformed by the potential.
    
    To test the degree to which the capture rate and Sommerfeld enhancement are related, we compare the numerically calculated radiative capture rate to the Sommerfeld enhancement, normalized by the ratio between these two quantities in the Coulomb regime:
    \begin{equation}
        \sigma v_{\text{rel}}=\left(\sigma v_{\text{rel}}\right)_C\frac{S}{S_{\text{C}}}.
    \end{equation}
    
    Here $\left(\sigma v_{\text{rel}}\right)_C$ is the capture cross section for a particular initial-state partial wave in the Coulombic limit ($\epsilon_{\phi}=\epsilon_{\delta}=0$), while $S$ and $S_C$ are the Sommerfeld enhancement factors (for the same partial wave) in the general pseudo-Dirac case and the Coulombic limit, respectively. 
    
    For a $s$-wave initial state, there is a semi-analytic approximation for pseudo-Dirac dark matter \cite{slatyer2010sommerfeld}, but in practice we use the numerically computed value in the comparison for both $s$- and $p$-wave initial-state contributions. The Sommerfeld enhancement for $s$-wave case in the Coulombic limit is given by \cite{Cirelli:2007xd,ArkaniHamed:2008qn,slatyer2010sommerfeld}: \begin{equation}
        S_{\text{C}}=\frac{\pi/\epsilon_v}{1-e^{-\pi/\epsilon_v}} \rightarrow \frac{\pi}{\epsilon_v}, \, \epsilon_v \ll 1,
    \end{equation} whereas for the $p$-wave case we have \cite{Cassel:2009wt}:\begin{equation}
        S_{\text{C}}=\frac{\pi/\epsilon_v}{(1-e^{-\pi/\epsilon_v})}\left(1+\frac{1}{4\epsilon_v^2}\right) \rightarrow \frac{\pi}{4\epsilon_v^3}, \, \epsilon_v \ll 1.
    \end{equation} 
    
    The capture rate is also analytically computable in the Coulombic limit, and takes a particularly simple form in the limit of small $\epsilon_v$. Using the results from Ref.~\cite{Asadi:2016ybp} we have:
    
    \begin{itemize} 
    \item $s \rightarrow 2p$:
    
    \begin{align} A &\approx \frac{2^5\sqrt{2 \pi}}{3} \alpha^{-1/2} m_\chi^{-3/2} \Gamma(1 - i / 2 \epsilon_v) e^{\pi/(4 \epsilon_v)} e^{-4}, \nonumber \\
    (\sigma v_\text{rel})_{s \rightarrow p} & \approx \frac{\alpha_\text{rad}}{v_\text{rel}} \frac{ \alpha^2 }{m_\chi^2} \frac{2^{11} \pi^2}{3^2} e^{-8} \end{align}
    
    This is the cross section for an initial $\chi^0 \chi^0$ spin-singlet state, summed over the three $2p$ bound states ($m=0,\pm 1$). To obtain the contribution to the spin-averaged cross section, one would multiply this cross section by a factor of 1/4. We have also chosen $k=\alpha^2 m_\chi/16$, as appropriate for a $n=2$ state in the Coulomb limit.
    
    \item $p \rightarrow 1s$:
     \begin{align} A &\approx 2^4 \sqrt{\pi} \alpha^{-1/2} m_\chi^{-3/2} \Gamma(1 - i / 2 \epsilon_v) e^{\pi/(4 \epsilon_v)} e^{-2}, \nonumber \\
    (\sigma v_\text{rel})_{s \rightarrow p} & \approx \frac{\alpha_\text{rad}}{v_\text{rel}} \frac{\alpha^2 }{m_\chi^2} \frac{2^{10} \pi^2}{3} e^{-4} \end{align}
    
    This is again the cross section for an initial state of fixed spin, in this case the spin-triplet configuration. To obtain the contribution to the spin-averaged cross section, this result should thus be multiplied by 3/4. We have set $k=\alpha^2 m_\chi/4$, since the bound state has $n=1$ in this case.
    \end{itemize}
    
\begin{figure}[h]
    \centering
    \includegraphics[scale=1]{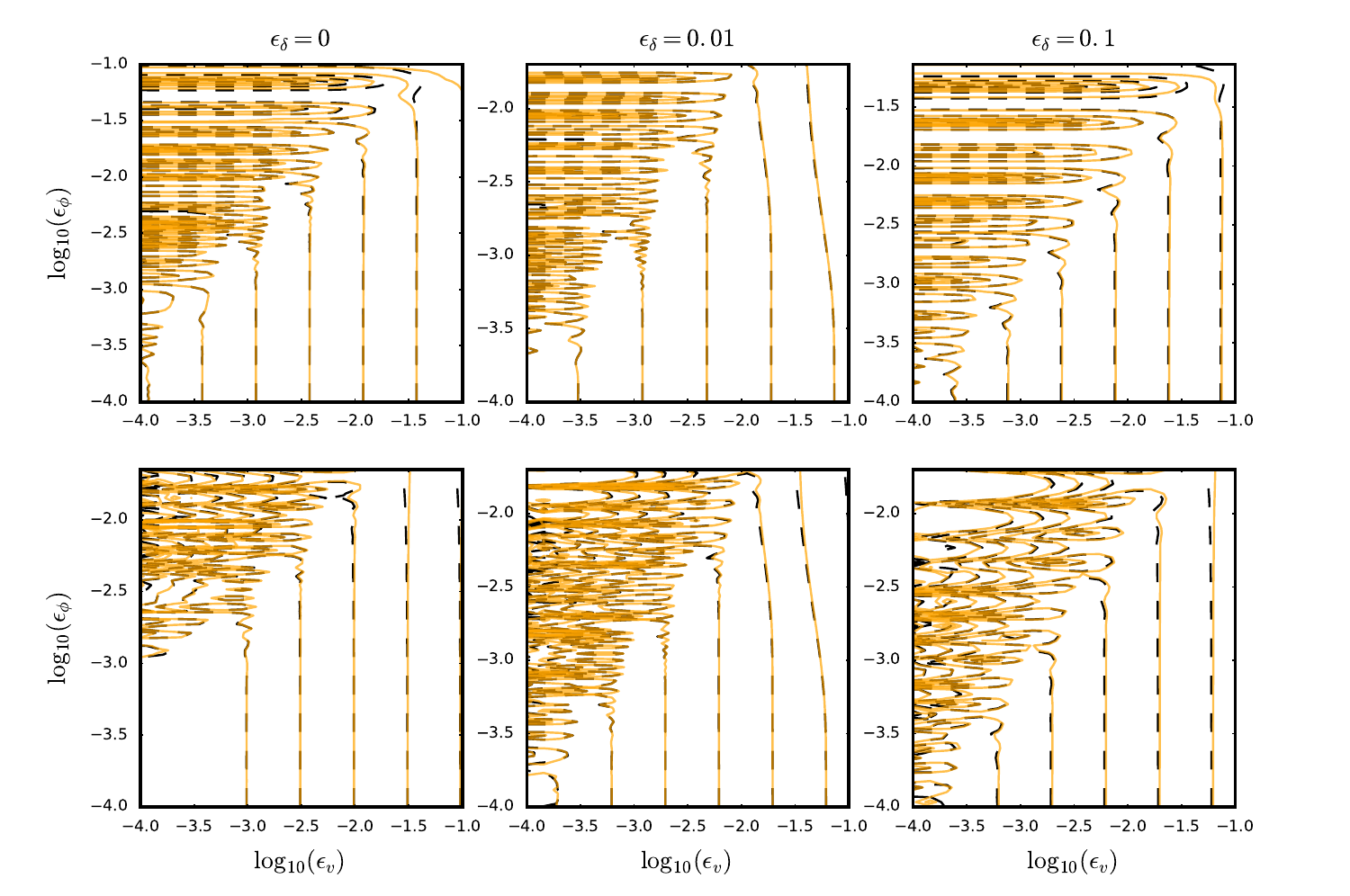}
    \caption{Comparison of the true capture cross section, $\log_{10}(\sigma v_{\text{rel}}m_{\chi}^2c/\hbar)$, and an estimate using the rescaled Sommerfeld enhancement,  $\log_{10}\left((\sigma v_{\text{rel}})_{\text{C}}m_{\chi}^2c/\hbar \frac{S}{S_{\text{C}}}\right)$, as a function of $\log_{10}(\epsilon_v)$ and $\log_{10}(\epsilon_{\phi})$. The solid orange lines are the numerical bound-state capture rate, and the dashed black lines are the rescaled Sommerfeld enhancement. The upper row indicates the $s\rightarrow 2p$ capture, while the lower is for the $p\rightarrow 1s$ transition. The columns are arranged in order of increasing mass splitting $\epsilon_{\delta}$ with the values being $0,0.01,0.1$ respectively. We find that they have similar behaviour, up to the phase space factor appearing in the bound-state formation rate.
    \label{fig:sommerfeld}}
\end{figure}

    As shown in Fig.~\ref{fig:sommerfeld}, the rescaled Sommerfeld enhancement has a very similar behaviour to the full bound-state formation rate. The principal difference between the two is captured in the phase-space factor in the bound state case (or rather, the ratio of the phase-space factor to its value in the Coulomb regime), which ensures that the bound state actually exists; this factor is simple and analytically calculable as soon as the bound-state energies are known. In the low-velocity limit for a massless emitted particle, we can approximate it by:
    \begin{equation}E/(\alpha^2 m_\chi/4 n^2) = 4 n^2 \epsilon_E \approx \left(1 - \frac{\pi^2}{6} n^2 \epsilon_\phi\right)^2 - 2 n^2 \epsilon_\delta^2.\end{equation}
    
    This result suggests that at least in this simple Abelian model and for capture into low-lying bound states, the ratio -- in a given partial wave -- of the bound-state formation rate to the Sommerfeld-enhanced annihilation rate is nearly independent of the parameters $\epsilon_\phi$ and $\epsilon_\delta$, and can be quite well-described by the (analytically calculable) ratio for the Coulombic limit with $\epsilon_\phi,\epsilon_\delta \rightarrow 0$, up to the phase-space factor. Thus if the same partial wave dominates both the Sommerfeld-enhanced annihilation rate and the bound-state formation, whichever process has a larger rate in the Coulomb case will also dominate for the bulk of the parameter space with $\epsilon_\phi,\epsilon_\delta \ne 0$, at least in the regime we have studied where bound states exist and the Sommerfeld enhancement is large. This conclusion need not hold, however, if e.g. the Sommerfeld-enhanced annihilation is dominated by $s$-wave processes but the bound-state formation is dominated by $p$-wave initial states capturing into a $s$-wave initial state.

\section{Conclusion}
\label{sec:conclusion}

Using a simple model of pseudo-Dirac dark matter interacting with a light vector boson as a testbed, we have derived an analytic expression for the shift in bound-state energies when the bound state has constituents of slightly different masses. In the two-state case, with a mass splitting of $2 \delta$ between the available two-particle configurations, and bound states comprised equally of these configurations, the shift in the binding energies $E$ is given by:

\begin{equation}
    E(\epsilon_{\phi},\epsilon_{\delta})=E(\epsilon_{\phi},0)-\delta
\end{equation}where $E$ is the binding energy and $\epsilon_{\phi}$ and $\epsilon_{\delta}$ are as defined in Eq.~\ref{eq:dimensionlessparams}. Thus, the mass splitting can be simply accounted for by subtracting half the mass splitting from the binding energy in the degenerate (mass) case. The expression is approximate and holds only for small mass splitting and small force-carrier mass, and we expect the regime of validity to be given roughly by:\begin{equation}
    \frac{E}{m_{\chi}\alpha^2}\gg \max\left(\text{min}\left[\epsilon_{\phi}^2/\ln(\epsilon_\phi/\epsilon_\delta^2)^2,\epsilon_\phi^2\right],\ \epsilon_{\delta}^4\right). 
\end{equation}

More generally, if there are $N$ two-body states involved, and the mass splittings between the states are captured in the Hamiltonian for these states as a $N\times N$ constant matrix $\Delta H$, a bound state characterized by a (unit normalized) $N$-vector $\eta$ in the space of two-body states experiences an offset in its binding energy of $\eta^\dagger \Delta H \eta$. 

In the case of wino DM, $L+S$-even bound states are an admixture of $\ket{\chi^0 \chi^0}$ and $\ket{\chi^+ \chi^-}$ two-body states, with coefficients $1/\sqrt{3}$ and $\sqrt{2/3}$ respectively. Since the $\chi^+ \chi^-$ state is heavier by a mass splitting $2\delta$, the binding energies for these states are offset by $2 \delta \times 2/3 = (4/3)\delta$. Combining this estimate with an approximate solution for bound-state energies in the Hulth\'en potential, we obtain an analytic estimate for the energies of $L+S$-even wino bound states:
\begin{equation}
E_n=m_{\chi}\alpha_{W}^2\left(\frac{1}{n}-\frac{n\ m_{W}\pi^2}{12\alpha_W m_{\chi}}\right)^2-\frac{4}{3}\delta
\end{equation} where $n$ is the principal quantum number of the bound state, $m_W$ is the mass of the $W^{\pm}$ boson, $\alpha_W$ is the electroweak coupling constant, and $\delta$ is the chargino-neutralino mass splitting between $\chi^0$ and $\chi^-$. Our estimate is in excellent agreement with previous numerical calculations \cite{Asadi:2016ybp}. 

This analytic expression for the binding energies enables us to estimate the locations of resonances in the scattering cross section due to near-zero energy bound states. For our model of pseudo-Dirac DM, we find that the resonances in the $\left(\epsilon_{\phi},\ \epsilon_{\delta}\right)$ plane obey the linear relationship \begin{equation}\label{eq:resshift}
    \epsilon_\phi = \frac{1}{n^2 q} \left(1 - \sqrt{2} n \epsilon_\delta \right)
\end{equation}where the numeric factor $q$ is approximately $\pi^2/6$ for the $s$  states, and is roughly of the same order for the higher $l$ states. This extends the observation of linear resonance shifts induced by a mass splitting, for the case of $l=0$, made in Ref.~\cite{slatyer2010sommerfeld}. 

We also analyzed the effect of the mass splitting on the cross section for radiative capture into bound states in our model of pseudo-Dirac DM. We find that in this case, the effects of the mass splitting on the capture rate are essentially identical to its effects on the Sommerfeld enhancement (for the same partial wave), up to phase space factors that depend on the mass of the emitted particle and the energy of the bound state being formed. Consequently, numerical calculations and analytic estimates for the Sommerfeld enhancement can be equally applied to estimate the bound-state formation rate. Furthermore, it is not feasible to significantly enhance bound-state formation relative to Sommerfeld-enhanced annihilation by turning on a mass splitting, unless different partial waves dominate the Sommerfeld enhancement and the bound-state formation rate.

The features inherited from the Sommerfeld enhancement include:
\begin{itemize}
    \item The resonances are enhanced relative to the zero mass splitting case, and undergo a shift at low velocities prescribed by Eq.~\ref{eq:resshift}.
    \item As a consequence of the shift in resonances, the bound-state formation rate at small velocities can develop a non-monotonic velocity dependence, even if it was monotonic with velocity for zero mass splitting.
    \item There is an ``adiabatic'' regime at non-zero mass splitting, where under appropriate values of the force carrier mass and relative velocity (as dictated by Eq.~\ref{eq:adiabatic_criterion}) the cross section is doubled relative to the corresponding zero-mass-splitting case. This is a generic feature of such multi-state systems, although the size of the enhancement varies (for example, it is a factor of three for wino DM).
 
\end{itemize}

\acknowledgments We thank Pouya Asadi, Rakhi Mahbubani, Gregory Ridgway and Chih-Liang Wu for helpful comments and discussions. This work was supported by the Office of High Energy Physics of the U.S. Department of Energy under grant Contract Numbers DE-SC0012567 and DE-SC0013999. TRS is partially supported by a John N. Bahcall Fellowship. TRS thanks the Galileo Galilei Institute for Theoretical Physics and the Kavli Institute for Theoretical Physics for hospitality during the completion of this work, and acknowledges partial support from the INFN, the Simons Foundation (341344, LA), and the National Science Foundation under Grant No. NSF PHY-1748958.

\bibliographystyle{unsrt}
\bibliography{ref.bib}

\end{document}